\documentclass{article}
\usepackage{amsmath,graphicx,amscd,siunitx,booktabs,subcaption,xurl}
\title{Sensitivity analysis of an epidemic model with a mass vaccination program of a homogeneous population}
\author{Ma.~Cristina~R.~Bargo}
\date{August 25, 2025}

\begin{document}

\maketitle

\begin{abstract}
The COVID-19 pandemic forced the rapid development of vaccines and the implementation of mass vaccination programs around the world. However, many hesitated to take the vaccine due to concerns about its effectiveness. By looking at an ordinary differential equation (ODE) model of disease spread that incorporates a mass vaccination program, this study aims to determine the sensitivity of the cumulative count of infected individuals ($W$) and the cumulative death count ($D$) to the following model parameters: disease transmission rate ($\beta$), reciprocal of the disease latency period ($\kappa$), reciprocal of the infectious period ($\gamma$), death ratio ($\alpha$), vaccine efficacy rate ($r$), and vaccine rollout rate ($\delta$). This was implemented using Latin hypercube sampling and partial rank correlation coefficient. Results show that $D$ is highly sensitive to $\alpha$ and shows increasing sensitivity to $\delta$ in the long run. On the other hand, $W$ is highly sensitive to $\kappa$ at the beginning of the simulation, but this weakens over time. In contrast, $W$ is not very sensitive to $\delta$ initially but becomes very significant in the long run.  This supports the importance of the vaccine rollout rate over the vaccine efficacy rate in curbing the spread of the disease in the population. It is also worthwhile to reduce the death ratio by developing a cure for the disease or improving the healthcare system as a whole. \\

\end{abstract}

\section{Introduction}
The COVID-19 pandemic that originated in Wuhan, China had a lasting impact on our society. Governments around the world were forced to implement lockdowns in an attempt to curb the spread of the pandemic, which proved detrimental to the economy \cite{adb}. Vaccines against COVID-19 were developed at an extraordinary rate, and massive vaccination programs were launched in many countries. 

Vaccine efficacy is a measure of how much a vaccine lowers the risk of getting the outcome being considered (for example, being infected by the disease) \cite{efficacy}. The World Health Organization only approves vaccines that have at least 50\% efficacy. The best vaccines were Moderna (89.2\% efficacy against symptomatic infections and 95\% efficacy against hospitalization) \cite{moderna} and Pfizer (91\% efficacy)\cite{pfizer}. On the other hand, China's Sinovac-CoronaVac had an efficacy of 51\% against symptomatic infections and 100\% efficacy against hospitalization \cite{sinovac}.

It was recently revealed in a Reuters investigation \cite{pentagon} that the Pentagon launched an anti-vaccination campaign to discredit Sinovac, the COVID-19 vaccine developed in China. These efforts were done via social media and targeted the Philippine population, eventually expanding to Southeast Asia, Central Asia, and the Middle East. The Philippines might have been chosen as a target because Filipinos are very active in social media \cite{socialmediastats}. Together with the Dengvaxia controversy in 2016, it is no surprise that a lot of Filipinos were hesitant to get the COVID-19 vaccine at the beginning of the vaccination program \cite{dengvaxia}. One study \cite{hesitancy} identified brand preferences, negative experiences with the health system, misinformation, and political issues as some of the contributing factors to vaccine hesitancy.

This study aims to explore the validity of brand preferences as a reason for vaccine hesitancy. We consider a slight variation of an epidemic model that incorporates an ongoing mass vaccination program \cite{model}. We use Latin hypercube sampling (LHS) and partial rank correlation coefficient (PRCC) to perform sensitivity analysis \cite{marino}. Our goal is to determine the sensitivity of the cumulative count of infected individuals and the death count to the vaccine efficacy and the rollout rate (or the rate of distribution) of the vaccine to the population, as well as the other disease parameters of the model. This study is contextualized in the Philippine setting, using the estimated model parameters from various sources.

Using the results from this study could provide insights into the implementation of mass vaccination programs in the event that another pandemic takes place in the future.  Proving the importance of vaccine efficacy would mean that WHO should be more stringent in approving vaccines that can be distributed to the population. However, proving the importance of vaccine rollout rate would mean that governments should figure out a way to alleviate the hesitancy of the population to vaccines.

\section{Materials and Methods}
This section covers the modified SEIR model that incorporates the implementation of a mass vaccination program. We also discuss the numerical solution of the model and determine the input and output variables that will be used for sensitivity analysis. Finally, we discuss the implementation of Latin hypercube sampling and partial rank correlation coefficient to determine the sensitivity of our chosen output variables with the model parameters.

\subsection{Model Formulation}
\begin{figure}
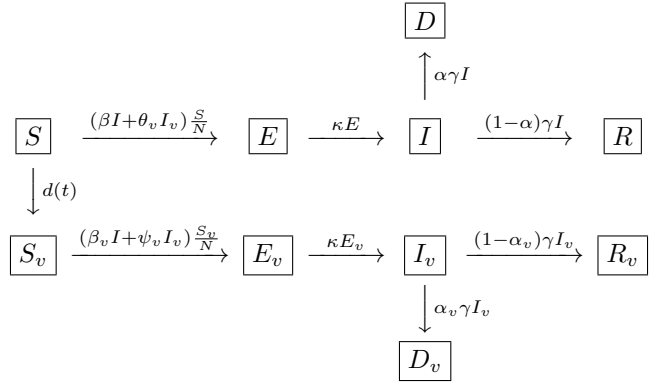

    $$\begin{CD}
	@. @. \boxed{D} @. \\
	@. @. @AA\alpha \gamma I A @.\\
	\boxed{S} @>(\beta I+\theta_v I_v)\frac{S}{N}>> \boxed{E} @>\kappa E>> \boxed{I} @>(1-\alpha) \gamma I>> \boxed{R}\\
	@VV d(t) V @. @. @.\\
	\boxed{S_v} @>(\beta_v I+\psi_v I_v)\frac{S_v}{N}>> \boxed{E_v} @>\kappa E_v>> \boxed{I_v} @>(1-\alpha_v)\gamma I_v>> \boxed{R_v} \\
	@. @. @VV\alpha_v \gamma I_v V @.\\
	@. @. \boxed{D_v} @.
\end{CD}$$
    \caption{Compartmental diagram of the SEIR model incorporating a mass vaccination program.}
    \label{fig:compartments}
\end{figure}

We consider a modified SEIR model that incorporates a mass vaccination program \cite{model}.  We define the following variables:
\begin{itemize}
	\item $S$: number of unvaccinated susceptible individuals
	\item $S_v$: number of vaccinated susceptible individuals
	\item $E$: number of unvaccinated exposed individuals
	\item $E_v$: number of vaccinated exposed individuals
	\item $I$: number of unvaccinated infected individuals
	\item $I_v$: number of vaccinated infected individuals
	\item $R$: number of unvaccinated recovered individuals
	\item $R_v$: number of vaccinated recovered individuals
	\item $D$: number of unvaccinated individuals who died from the disease
	\item $D_v$: number of vaccinated individuals who died from the disease
    \item $N$: total initial population.
\end{itemize}
Refer to Figure \ref{fig:compartments} for the compartmental diagram of this model.
We make the following assumptions on our model.
\begin{enumerate}
	\item Vaccines are administered only to susceptible individuals. Susceptibles are being vaccinated at the rate $d(t)$. To prevent $S$ from becoming negative, we define
	\begin{equation}
		d(t) = \mathrm{min} \left\{ \delta, S^\star (t) \right\},
	\label{eq:rolloutrate}
	\end{equation}
	where $\delta$ is the vaccination rate at the beginning of the vaccination program and $S^\star (t)$ is the number of remaining qualified susceptibles at time $t$, i.e., the susceptibles who won't be transferred to the unvaccinated exposed class in the next time step \cite{teodoro}. At the beginning of the vaccination program, the vaccines are being distributed at a constant rate $\delta$ that is dictated by the supply of vaccines and the manpower of the organization that conducts the vaccination program. However, once the current number of susceptibles $S(t)$ falls below $\delta$, then only the remaining qualified susceptibles $S^\star (t)$ will be vaccinated.
	
	\item The vaccine is imperfect in the sense that it does not completely prevent an individual from contracting the disease. However, a vaccinated susceptible individual is less likely to be infected. If ever a vaccinated individual is infected, the viral load is less than the viral load of an unvaccinated individual.
	
	Let $\beta$ be the disease transmission rate when an unvaccinated susceptible individual interacts with an unvaccinated infected individual.  Let $\theta_v$ be the disease transmission rate when an unvaccinated susceptible interacts with a vaccinated infected individual. Let $\beta_v$ be the disease transmission rate when a vaccinated susceptible interacts with an unvaccinated infected individual. Let $\psi_v$ be the disease transmission rate when a vaccinated susceptible interacts with a vaccinated infected individual.  Because the vaccine is imperfect, we have $0 < \theta_v, \beta_v, \psi_v < \beta$. In this work, we adopt the following parameters \cite{model}:
	\begin{align}
		\theta_v &= \dfrac{1}{3} \beta \label{eq:thetav} \\
		\beta_v &= (1-r) \beta \label{eq:betav}\\
		\psi_v &= \dfrac{1}{3} \beta_v \label{eq:psiv}
	\end{align}
	where $r$ is the efficacy of the vaccine distributed in the population.
	
	\item The latency period $1/\kappa$ is defined to be the period from the moment that a susceptible acquires the infection up to the moment when the symptoms show up and the individual is now classified as an infected. The rate of transfer of individuals from the exposed compartments ($E$ and $E_v$) to the infected compartments ($I$ and $I_v$) is inversely proportional to the latency period, and hence is equal to $\kappa$. We assume that this rate is the same for both unvaccinated and vaccinated individuals.
    
    \item The infectious period $1/\gamma$ refers to the duration of the disease from the moment that the symptoms show up until the moment that the individual either recovers or dies from the disease. The rate of transfer of individuals from the infected compartments ($I$ and $I_v$) is inversely proportional to the infectious period, and hence is equal to $\gamma$. We assume that this rate is the same for both unvaccinated and vaccinated individuals.

    \item No reinfection occurs, which means a recovered individual gains immunity from the disease and does not go back to being susceptible.
	
	\item A certain proportion of infected individuals may die from the disease. This ratio is equal to $\alpha$ for unvaccinated individuals, and equal to $\alpha_v$ for vaccinated individuals.
		
	\item No additional individuals are added to the population (via births or immigration) or removed from the population (via emigration or death by other means). As a result, we have
	\begin{equation*}
	S + S_v + E + E_v + I + I_v + R + R_v + D + D_v = N
	\end{equation*}
	at any time $t$.
\end{enumerate}
The modified SEIR model that incorporates a mass vaccination program, under the assumptions mentioned above, can be described by this system of ordinary differential equations (ODEs):
\begin{equation}
	\begin{array}{rcl}
		S^\prime &=& -(\beta I+\theta_v I_v)\frac{S}{N} - \delta (t) \\
		S_v^\prime &=& -(\beta_v I+\psi_v I_v)\frac{S_v}{N} + \delta (t)\\
		E^\prime &=& (\beta I+\theta_v I_v)\frac{S}{N} - \kappa E\\
		E_v^\prime &=& (\beta_v I+\psi_v I_v)\frac{S_v}{N} - \kappa E_v \\
		I^\prime &=& \kappa E - \gamma I \\
		I_v^\prime &=& \kappa E_v - \gamma I_v \\
		R^\prime &=& (1-\alpha) \gamma I \\
		R_v^\prime &=& (1-\alpha_v) \gamma I_v \\
		D^\prime &=& \alpha \gamma I \\
		D_v^\prime &=& \alpha_v \gamma I
	\end{array}
	\label{eq:one-city}
\end{equation}

\subsection{Input and Output Variables}
Numerical simulations were implemented in Matlab R2024b. We solve the system (\ref{eq:one-city}) on the time interval $t \in [0,140]$ and discretize the interval with $\Delta t = 1$, indicating that we are monitoring daily counts of each compartment for 20 weeks. We choose to simulate over a short period to mimic the conditions of the mass distribution of vaccines during the COVID-19 pandemic, where supplies are limited because the vaccines were newly-developed.

We are going to fix the following parameters for the sensitivity analysis on model (\ref{eq:one-city}):
\begin{itemize}
	\item The total initial population of our city is $N = \num{5000000}$.
	\item The number of unvaccinated infected individuals at $t=0$ is $I(0) = \num{5000}$.  The rest of the population are unvaccinated susceptibles at $t=0$, so $S(0) = \num{4995000}$. The other compartments ($E$, $R$, $D$, $S_v$, $E_v$, $I_v$, $R_v$, $D_v$) are set to 0 at $t=0$.
    \item The vaccine prevents death from the disease, so $\alpha_v = 0$.
\end{itemize}
The remaining variables will be used as input variables in the sensitivity analysis in this study. These include:
\begin{itemize}
    \item disease parameters: transmission rate ($\beta$), reciprocal of the latency period ($\kappa$), reciprocal of the infectious period ($\gamma$), death ratio ($\alpha$)
    \item vaccine parameters: vaccine efficacy rate ($r$), vaccine rollout rate ($\delta$)
\end{itemize}

We solve the systems numerically using the explicit Euler scheme \cite{NumericalAnalysis}. While it is not as accurate as other numerical schemes in solving ODE systems, it allows us to find a nice expression for the number of susceptibles $S^\star(t)$ in equation (\ref{eq:rolloutrate}) that are qualified to receive the vaccine once the number of remaining susceptibles fall below the vaccine rollout rate. This means
\begin{equation}
	S^\star (t)  = S(t) - \left(\beta I(t) + \theta_v I_v(t)\right)\frac{S(t)}{N}.
\end{equation}

At the beginning of the COVID-19 pandemic, countries monitored cases by examining the cumulative number of infected individuals and the cumulative death count \cite{covidgraphs}. Hence, we shall choose the following model outputs for our analysis:
\begin{itemize}
	\item the cumulative death count at weekly itervals $t=7,14, \dots,140$, and
	\item cumulative infected count at weekly intervals $t=7,14,\dots,140$. 
\end{itemize}
The cumulative death count is given by $D(t)$, since the model assumes that the vaccine provides protection from death due to the disease ($\alpha_v=0$). On the other hand, let $W(t)$ be the cumulative infected count at time $t$. This satisfies the differential equation
\begin{equation}
	W^\prime = \kappa (E + E_v),
\label{eq:cuminf1}
\end{equation}
since we are counting the individuals who got infected regardless of their vaccination status. The initial condition is given by $W(0) = I(0) + I_v(0)$.

\begin{figure}
	\centering
	\includegraphics[width=0.75\textwidth]{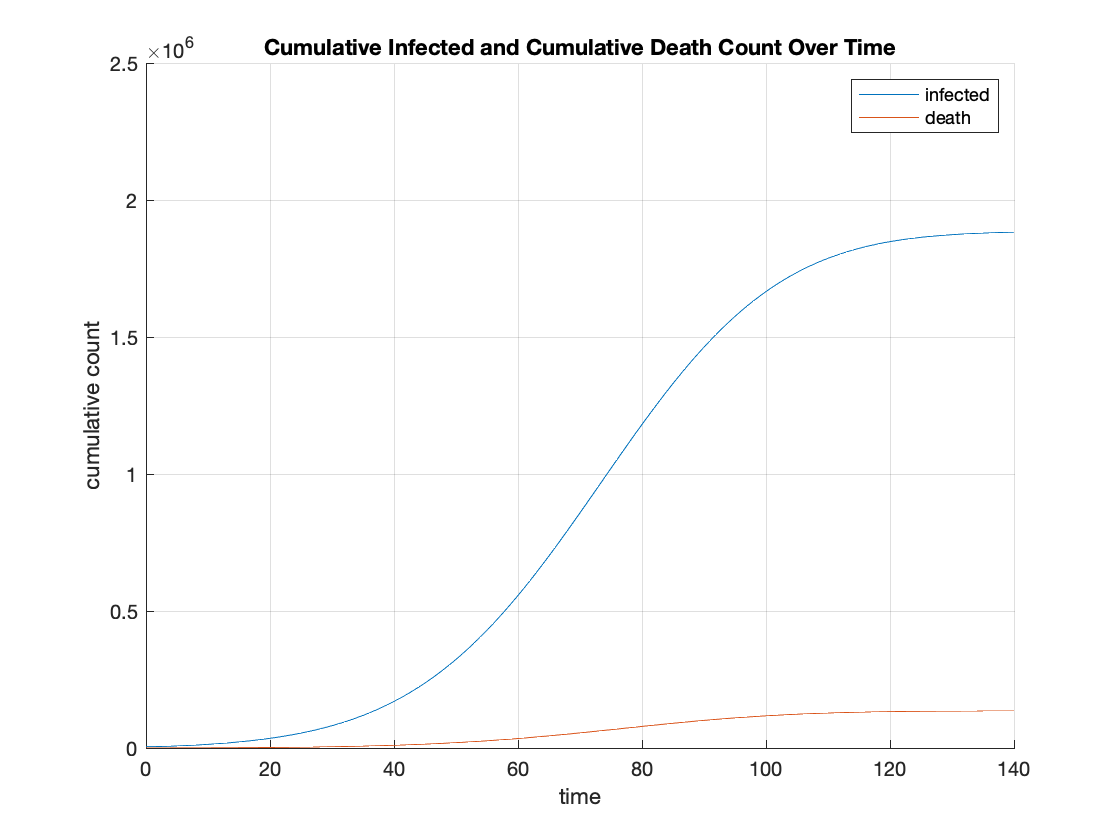}
	\caption{Evolution of the cumulative infected count $W$ (in blue) and the cumulative death count $D$ (in red) for the following parameters: $\beta = 0.37$, $\kappa = 0.18$, $\gamma = 0.15$, $\alpha = 0.09$,  $r=0.75$, and $\delta = \num{27887}$.}
	\label{fig:modelsim}
\end{figure}

Refer to Figure \ref{fig:modelsim} for the simulation of model (\ref{eq:one-city}) for the following values of the parameters: $\beta = 0.37$, $\kappa = 0.18$, $\gamma = 0.15$, $\alpha = 0.09$,  $r=0.75$, and $\delta = \num{27887}$. These values will serve as base values for the sensitivity analysis (to be discussed in the next section).
We are only interested in the graphs of the cumulative infected count over time $W(t)$ and the cumulative death count over time $D(t)$ to see the behavior of our chosen model output variables. We observe an increase in the cumulative infected and cumulative death count at around $t=60$, after which we observe that the graph approaches a horizontal asymptote.

\subsection{Sensitivity Analysis}
Sensitivity analysis aims to identify which of the input variables are the main contributors to the variation of the model output \cite{imanhelton}. A popular method of doing sensitivity analysis uses the combination of LHS and PRCC. This has been demonstrated on an HIV model \cite{blower}.

LHS is a variation of the stratified random sampling method that ensures each input variable of the model has all portions of the probability distribution represented \cite{mckay}. This involves dividing the range of each input variable into $N$ subintervals of equal probability, and then a sample is picked from each subinterval. The choice of $N$ should be at least equal to $k+1$ (where $k$ is the number of input variables of the model), but in practice $N > (4k)/3$ is a good criterion for choosing $N$ \cite{blower}.  We then generate the LHS matrix $X$ by taking the samples picked from the first input variable and pairing them randomly with the samples picked from the second input variable. These pairs are now assigned randomly to the samples picked from the third input variable, and so on, until all input variables have been assigned.  This produces a matrix of size $N \times k$, where each row contains a particular combination of the input variables for our model. We run the model for each row of the LHS matrix $X$ to obtain the values of the model output and store them on the output vector $Y$.

PRCC starts with rank-transforming the data from the LHS matrix and the output vector.  We then generate linear regression models described by
\begin{align}
	\hat{x}_j &= c_0 + \sum_{\substack{p=1 \\ p \neq j}}^k c_p x_p \\
	\hat{y} &= b_0 + \sum_{\substack{p=1 \\ p \neq j}}^k b_p x_p
\end{align}
for each input variable.  We then compute the correlation coefficient between the residuals $(x_j - \hat{x}_j)$ and $(y - \hat{y})$. We also compute the associated $p$-value for the correlation, defined to be the probability of observing a nonzero correlation coefficient with the assumption that the null hypothesis (i.e. there is no correlation between the parameter and the model output) is true. If the $p$-value is less than 0.05, we reject the null hypothesis and conclude that the correlation coefficient is significantly different from 0.

To implement sensitivity analysis on model (\ref{eq:one-city}), we gather information on the probability distributions of our chosen input variables. We focus our study with parameter estimates using Philippine data, summarized in Table \ref{tab:inputparams1}. 
\begin{table}
\centering
\caption{Input variables, baseline values and distribution for the model of disease spread with an ongoing mass vaccination program (\ref{eq:one-city}). }
\begin{tabular}{llll}
	\toprule
	Variable & Base Value & Distribution & Parameters \\
	\midrule
	$\mathcal{R}_0$ & $2.41$ & truncated normal & $\mu=2.41$ \\
     & & & $\sigma = 0.03826531$ \\
     & & & Range: $[0,+\infty)$ \\
	$1/\kappa$ & $5.5$ & truncated normal & $\mu = 5.5$ \\
     & & & $\sigma = 0.97$ \\
     & & & Range: $[1,+\infty)$ \\
	$1/\gamma$ & $6.5$ & truncated normal & $\mu = 6.5$ \\
     & & & $\sigma = 0.77$ \\
     & & & Range: $[4,+\infty)$ \\
	$\alpha$ & $0.09$ & uniform & Range: $[0,0.44]$ \\
	$r$ & $0.75$ & uniform & Range: $[0.5,1]$ \\
	$\delta$ & $\num{27887}$ & beta & $\alpha=0.7340199944487569$ \\
     & & & $\beta=112.5201094419424$ \\
     & & & $\textrm{scale} = \num{4302834.680007711}$ \\
     & & & $\textrm{offset} = -1.197495370221827 \times 10^{-23}$ \\
	\bottomrule
\end{tabular}
\label{tab:inputparams1}
\end{table}

We adopt the estimated probability distributions for the latency period $1/\kappa$ and the infectious period $1/\gamma$ from a study \cite{caldwell} using data from the Philippines from March 1, 2020, to February 23, 2021. We choose the base values of these parameters to be equal to the mean. For the transmission rate $\beta$, we use the distribution obtained by another study \cite{haw} for the basic reproduction number $\mathcal{R}_0$ (i.e. the average number of secondary infections from one infective individual) using data from the Philippines pre-ECQ phase (from February 11, 2020 to March 19, 2020), which removes the effect of social distancing protocols on the transmission of COVID-19. Similar to the case of the latency period and the infectious period, we choose the mean to be the base value for $\mathcal{R}_0$. Using the definition of the basic reproduction number $\mathcal{R}_0 = \beta/\gamma$, we then derive the transmission rate $\beta$ by dividing $\mathcal{R}_0$ by the base value of the infectious period $1/\gamma$ (equal to 6.5).

For estimating the death ratio $\alpha$, we refer to a study on the COVID-19 death tally as of April 29, 2020 \cite{haw}. The nationwide death ratio $\alpha$ is 0.09, but it varies based on geographic location, age group, and comorbidity. The highest death ratio is for the age group 80 and above (0.44). In the absence of studies on the probability distribution for the COVID-19 death ratio, we simply choose a uniform distribution in the range $[0,0.44]$ and set the nationwide death ratio of 0.09 to be the base value for $\alpha$. For the vaccine efficacy $r$, we also choose a uniform distribution because we have no information on the distribution of the efficacy of existing COVID-19 vaccines. We limit the range of the uniform distribution in the interval $[0.5,1]$ since the WHO only authorizes the distribution of vaccines with at least 50\% efficacy. We assign the middle of the interval (0.75) as the base value for $r$. As for the vaccine rollout $\delta$, data from the Philippines was obtained from Our World in Data \cite{vaccrollout} and the best probability distribution that fits the data was computed using the Python library \texttt{distfit} \cite{distfit}. Details of the calculation of probability distribution parameters are presented by Bargo \cite{deltacode}. The base value is equal to the mean of the beta distribution that fits the data.

To generate the LHS matrix, we have $k=6$ input variables and choose to divide the sample space into $N=700$ subintervals for each input variable. The Matlab implementation of LHS was based on the work of Minasny \cite{lhscode}, modified for the truncated normal distributions and the beta distribution. For the implementation of PRCC, we say that the correlation coefficient is significantly different from 0 when the $p$-value is less than 0.05. Since we are looking at the cumulative death count and the cumulative infected count at several time nodes, we shall be looking at the evolution of the correlation coefficients over time.

It should be noted that PRCC is not accurate when the relationship between the input variable and the model output is not monotonic, which means that the model output should be either increasing or decreasing as you increase the value of the input variable \cite{marino}.  To check for monotonicity of a model output with respect to an input variable, we take the corresponding column from the LHS matrix we generated previously. We then calculate the model output using the different values of the input variable while fixing the rest of the parameters using the base values in Table \ref{tab:inputparams1}. Although the cumulative infected count $W(t)$ remains constant with respect to the death ratio $\alpha$ for all time nodes, the rest of the parameters in Table \ref{tab:inputparams1} have been verified to have monotonic relationships with the output variables $D(t)$ and $W(t)$, $t=7,14,\dots,140$.

\section{Results and Discussion}
We present the evolution of the PRCC of the output variables (cumulative death count $D$ and cumulative infected count $W$) with respect to the parameters mentioned in Table \ref{tab:inputparams1}. We first present the results for the disease parameters: transmission rate ($\beta$), reciprocal of the latency period ($\kappa$), reciprocal of the infectious period ($\gamma$), and death ratio ($\alpha$). We then discuss the results for the vaccine parameters: vaccine efficacy rate ($r$) and vaccine rollout rate ($\delta$).

\subsection{Sensitivity with Respect to Disease Parameters}
\begin{figure}
	\centering
	\begin{subfigure}{0.475\textwidth}
		\includegraphics[width=\textwidth]{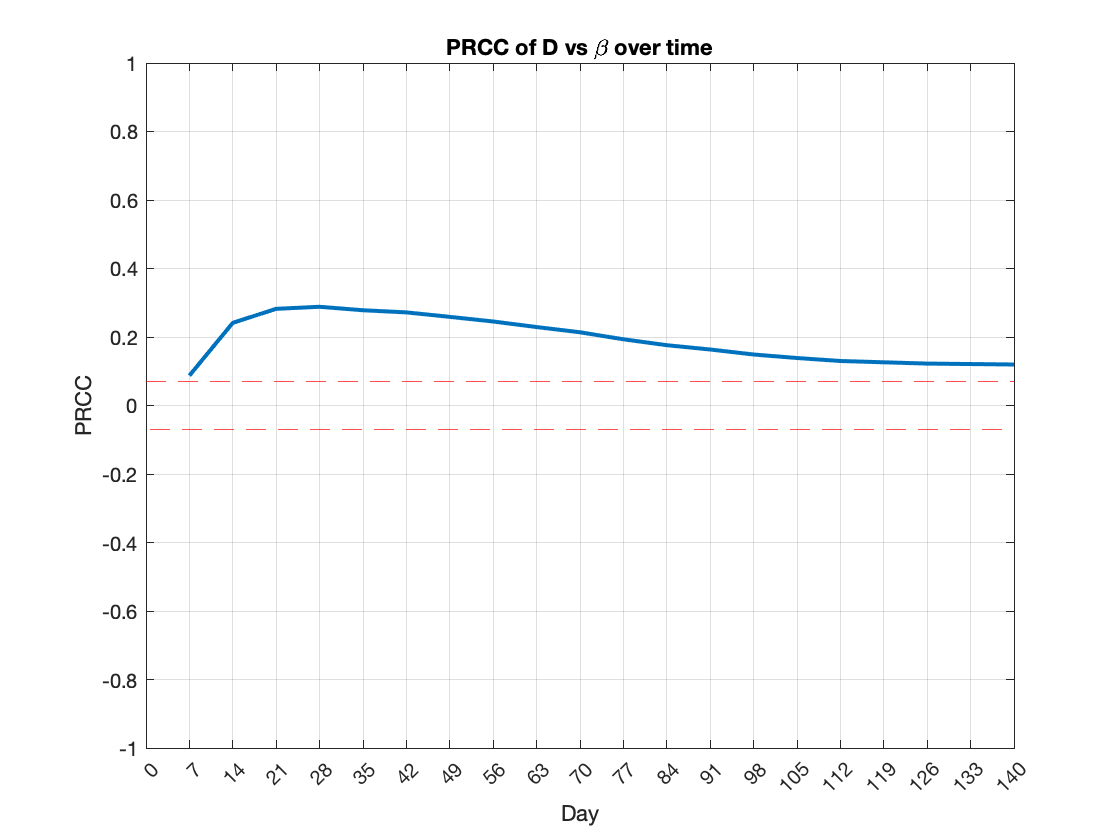}
		\caption{$D(t)$ vs $\beta$}
		\label{subfig:Dbeta}
	\end{subfigure}
	\begin{subfigure}{0.475\textwidth}
		\includegraphics[width=\textwidth]{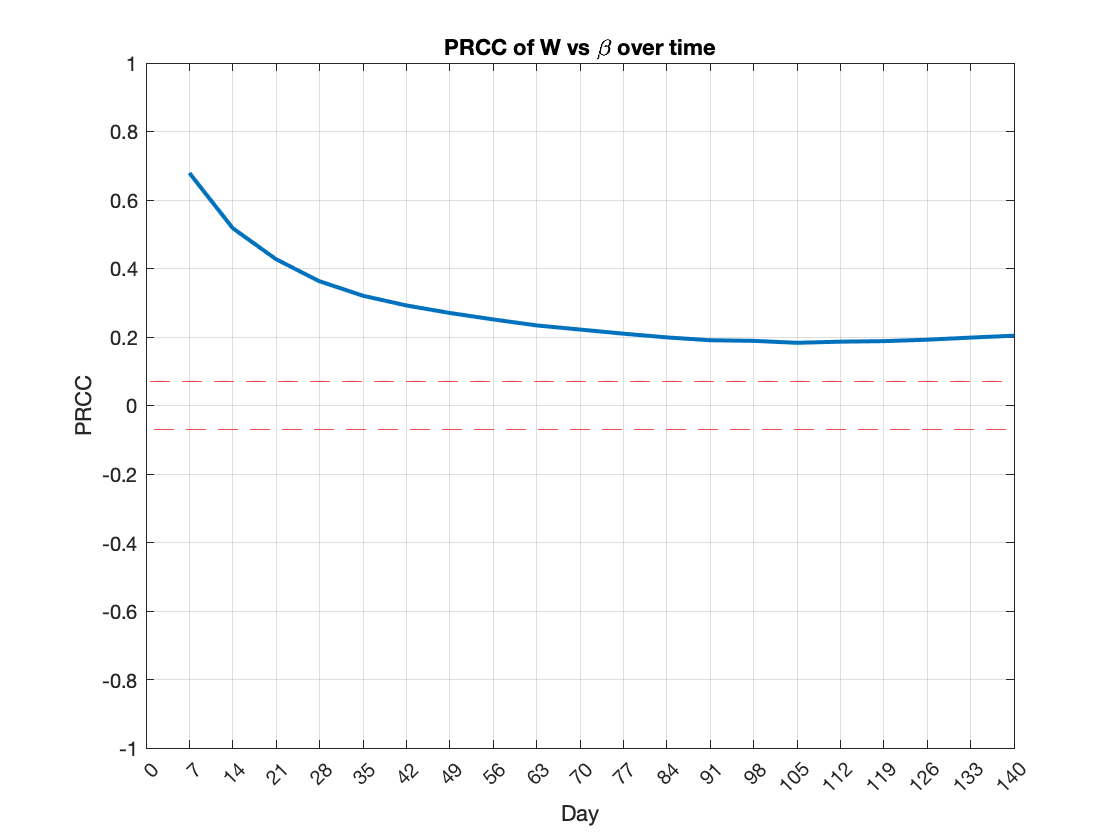}
		\caption{$W(t)$ vs $\beta$}
		\label{subfig:Wbeta}
	\end{subfigure}
	\caption{Evolution of the PRCC of the output parameters $D$ (Figure \ref{subfig:Dbeta}) and $W$ (Figure \ref{subfig:Wbeta}) with respect to $\beta$. The region between the two red lines indicate PRCC values that are not significantly different from 0 (where $p$-values are greater than 0.05), which is not the case for $\beta$.}
	\label{fig:beta}
\end{figure}

Figure \ref{fig:beta} shows the evolution of the PRCC of the output variables $D$ and $W$ with respect to $\beta$. Both output variables show a positive correlation, though we observe larger PRCC values for $W$ at the beginning of the simulation period (when we are at the beginning of the mass vaccination program). It is interesting to note that the PRCC values of $W$ versus $\beta$ decrease over time. On the other hand, we see an initial increase of the PRCC values for $D$ versus $\beta$ up to Day 28, and then it eventually decreases. The corresponding $p$-values are close to 0, indicating that the PRCC values are significantly different from 0.

\begin{figure}
	\centering
	\begin{subfigure}{0.475\textwidth}
		\includegraphics[width=\textwidth]{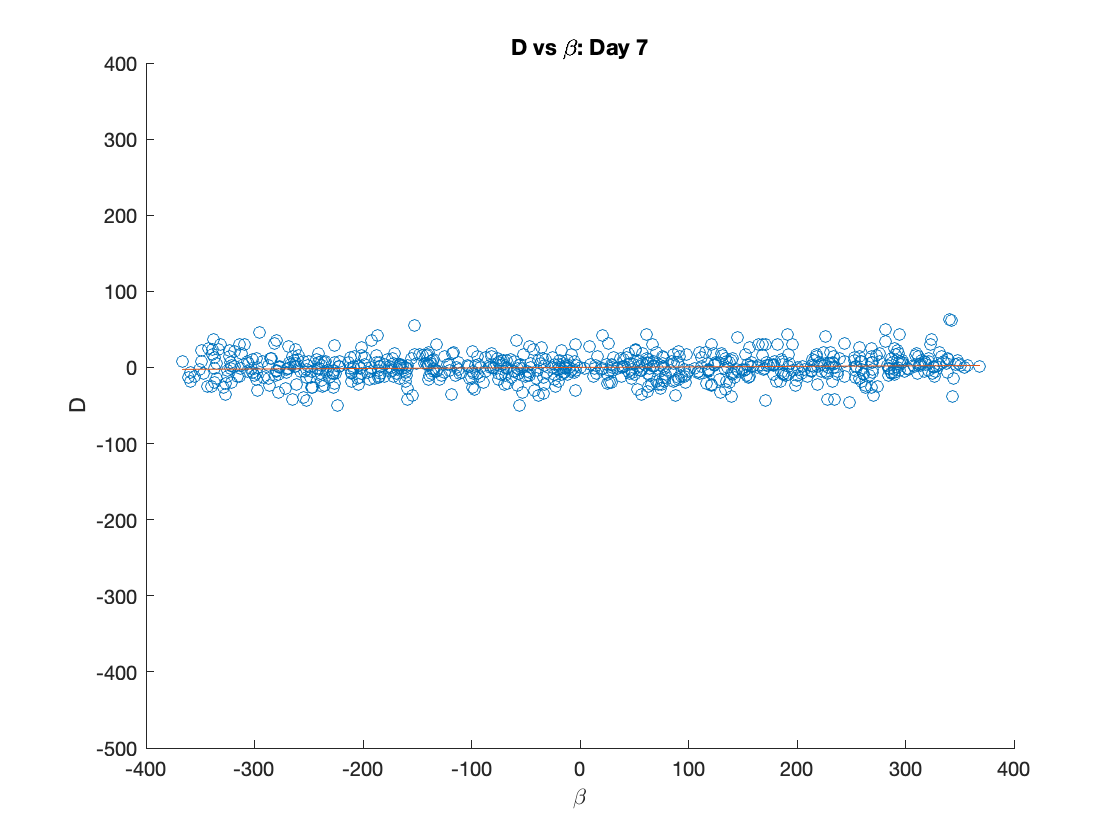}
		\caption{$D(t)$ vs $\beta$: Day 7}
		\label{subfig:Dbeta7}
	\end{subfigure}
	\begin{subfigure}{0.475\textwidth}
		\includegraphics[width=\textwidth]{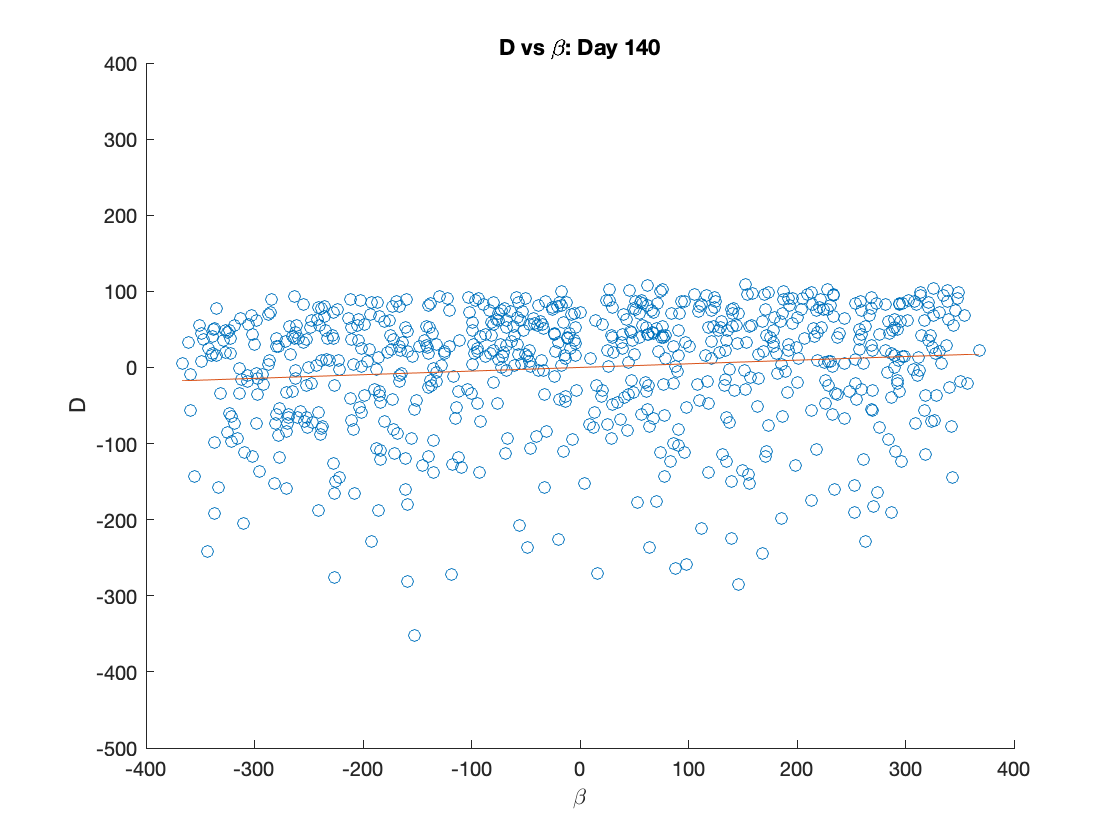}
		\caption{$D(t)$ vs $\beta$: Day 140}
		\label{subfig:Dbeta140}
	\end{subfigure}
	\begin{subfigure}{0.475\textwidth}
		\includegraphics[width=\textwidth]{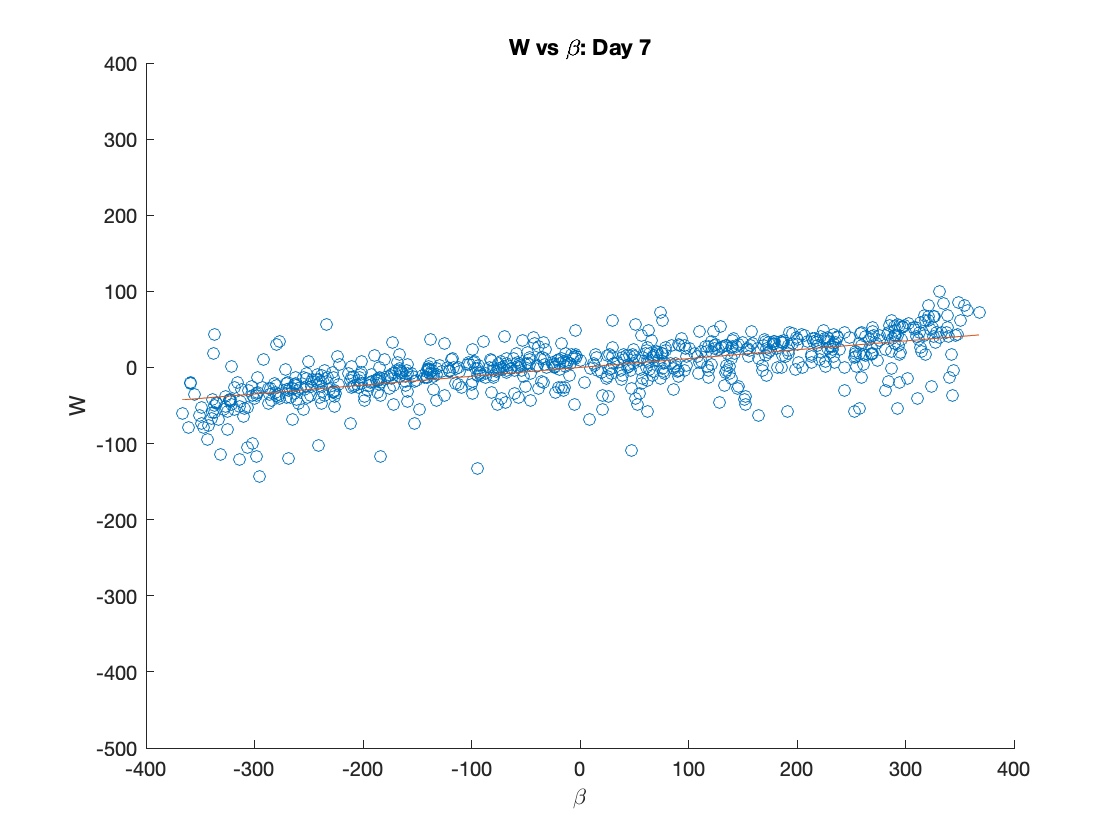}
		\caption{$W(t)$ vs $\beta$: Day 7}
		\label{subfig:Wbeta7}
	\end{subfigure}
	\begin{subfigure}{0.475\textwidth}
		\includegraphics[width=\textwidth]{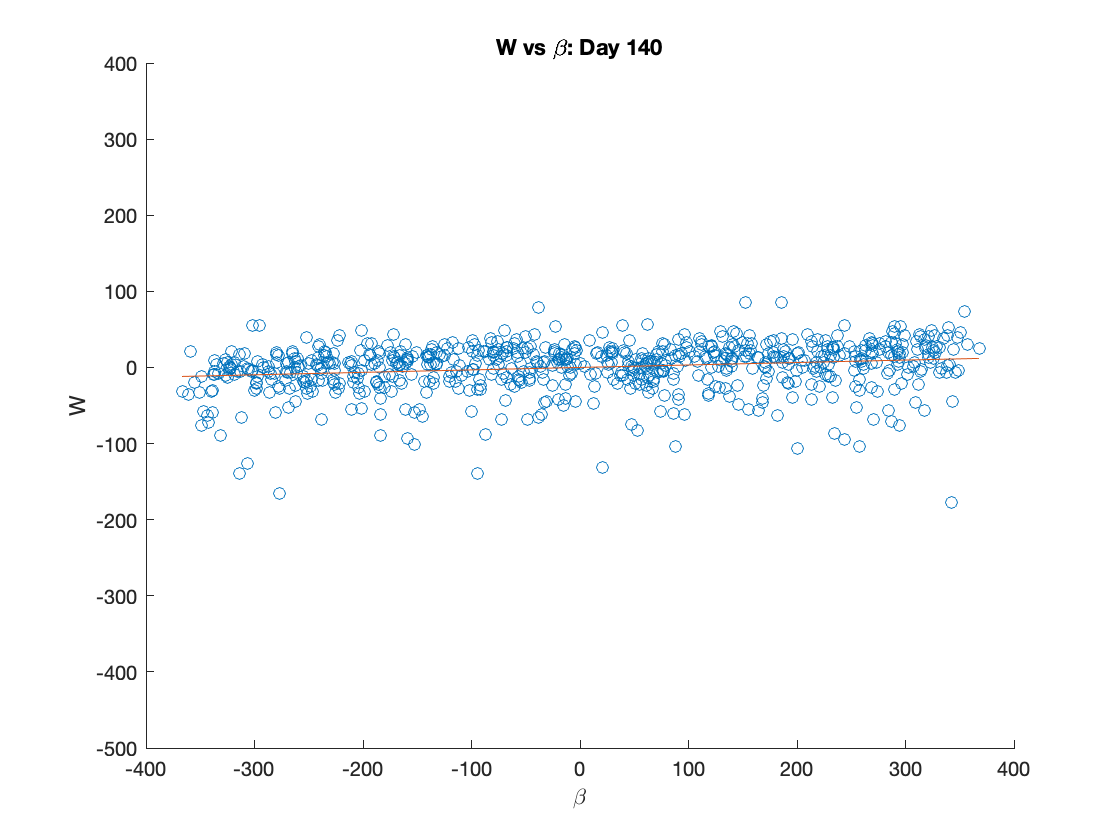}
		\caption{$W(t)$ vs $\beta$: Day 140}
		\label{subfig:Wbeta140}
	\end{subfigure}
	\caption{Scatter plots of the residuals of the rankings for $D$ (Figure \ref{subfig:Dbeta7} and \ref{subfig:Dbeta140}) and $W$ (Figure \ref{subfig:Wbeta7} and \ref{subfig:Wbeta140}) with respect to the residuals of the $\beta$ ranking.}
	\label{fig:betascatter}
\end{figure}

We also look at the scatter plots of the residuals of the output variables $D$ and $W$ with respect to $\beta$, as shown in Figure \ref{fig:betascatter}. The best-fit lines are almost horizontal, but we observe a better fit at Day 7. On Day 140, the points are more dispersed, with $D$ showing more dispersion compared with $W$. This is consistent with the plots in Figure \ref{fig:beta} showing that the PRCC values are getting closer to 0 towards the end of the simulation period. This suggests that both output variables are only slightly sensitive to $\beta$ and this sensitivity decreases over time. Efforts to reduce $\beta$ by reducing contact between susceptible and infected individuals (via social distancing rules) may not have a significant effect in reducing disease spread.

\begin{figure}
	\centering
	\begin{subfigure}{0.475\textwidth}
		\includegraphics[width=\textwidth]{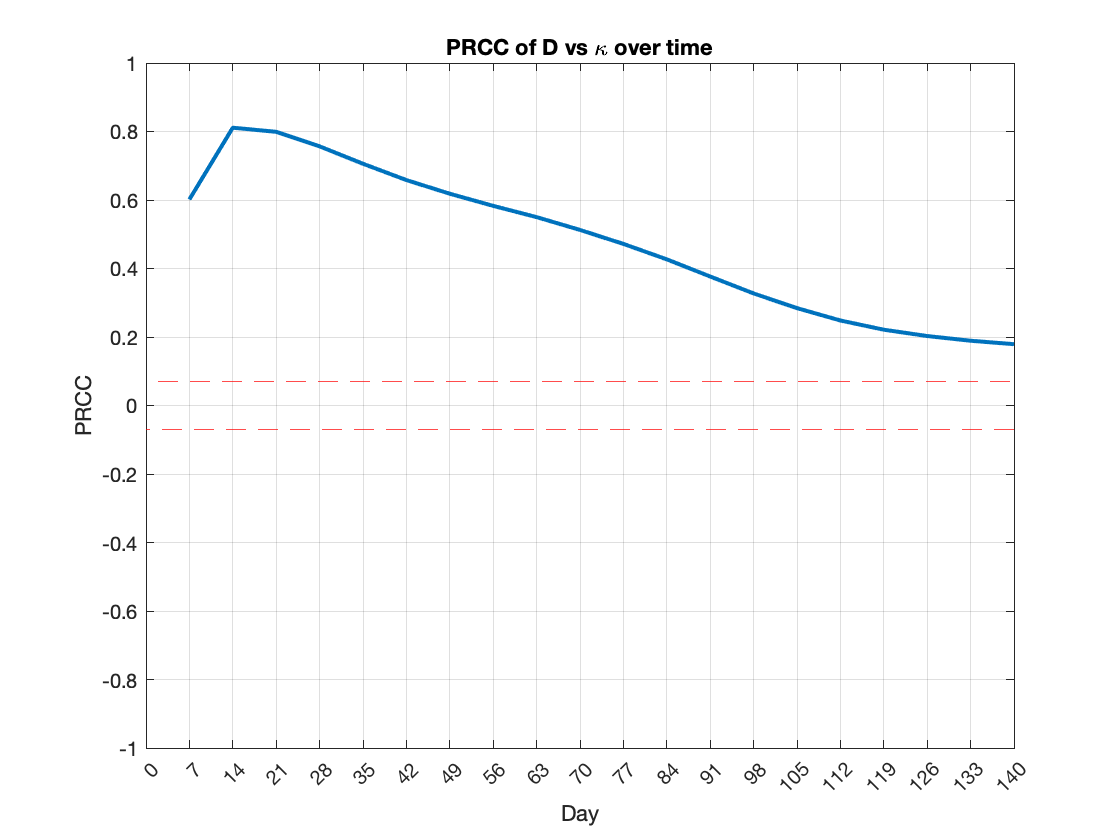}
		\caption{$D(t)$ vs $\kappa$}
		\label{subfig:Dkappa}
	\end{subfigure}
	\begin{subfigure}{0.475\textwidth}
		\includegraphics[width=\textwidth]{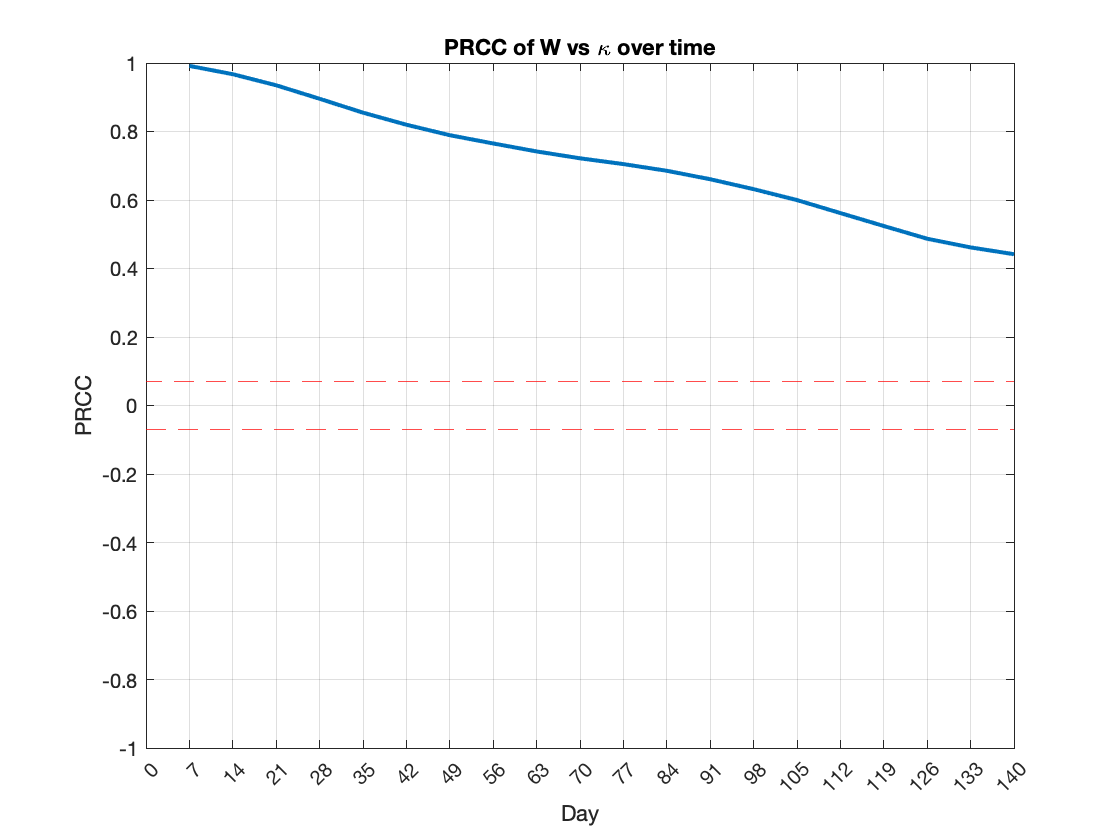}
		\caption{$W(t)$ vs $\kappa$}
		\label{subfig:Wkappa}
	\end{subfigure}
	\caption{Evolution of the PRCC of the output parameters $D$ (Figure \ref{subfig:Dkappa}) and $W$ (Figure \ref{subfig:Wkappa}) with respect to $\kappa$. The region between the two red lines indicate PRCC values that are not significantly different from 0 (where $p$-values are greater than 0.05), which is not the case for $\kappa$.}
	\label{fig:kappa}
\end{figure}

Figure \ref{fig:kappa} shows the evolution of the PRCC of the output variables $D$ and $W$ with respect to $\kappa$. Both output variables show a positive correlation, and we also observe this correlation decrease over time (similar to $\beta$). The corresponding $p$-values are very small, implying that the PRCC values are significantly different from 0.

\begin{figure}
	\centering
	\begin{subfigure}{0.475\textwidth}
		\includegraphics[width=\textwidth]{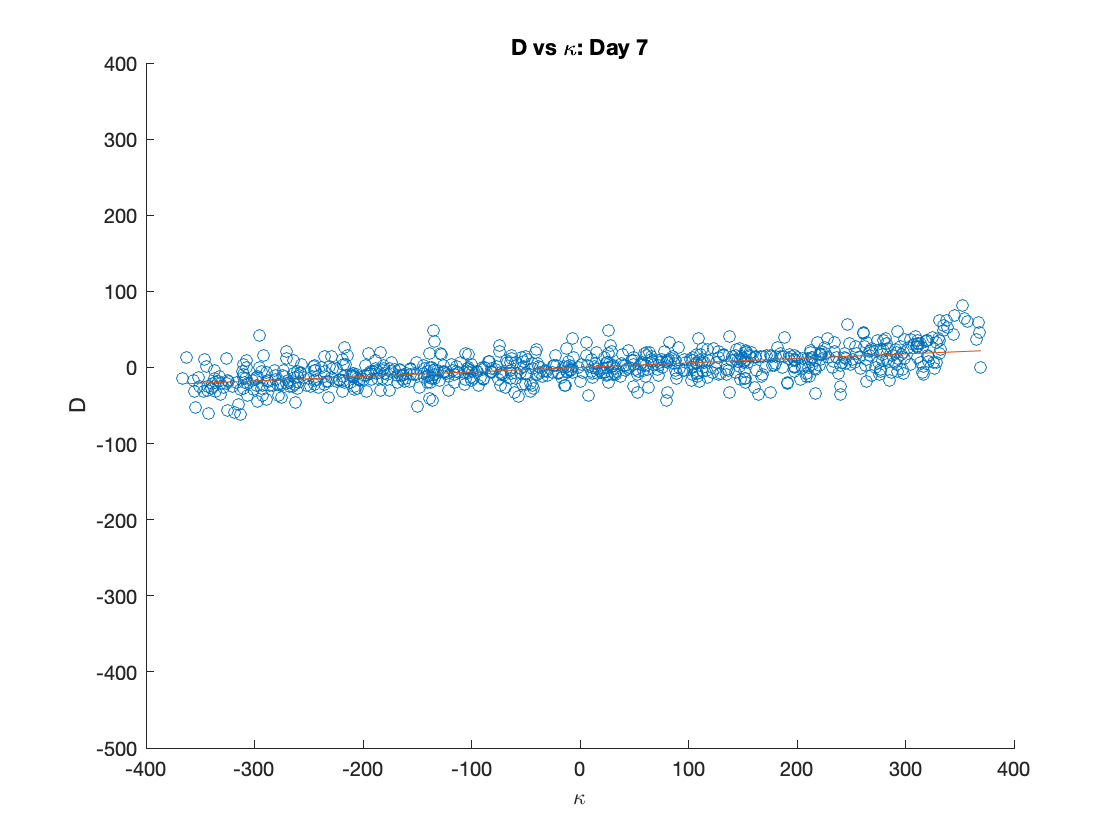}
		\caption{$D(t)$ vs $\kappa$: Day 7}
		\label{subfig:Dkappa7}
	\end{subfigure}
	\begin{subfigure}{0.475\textwidth}
		\includegraphics[width=\textwidth]{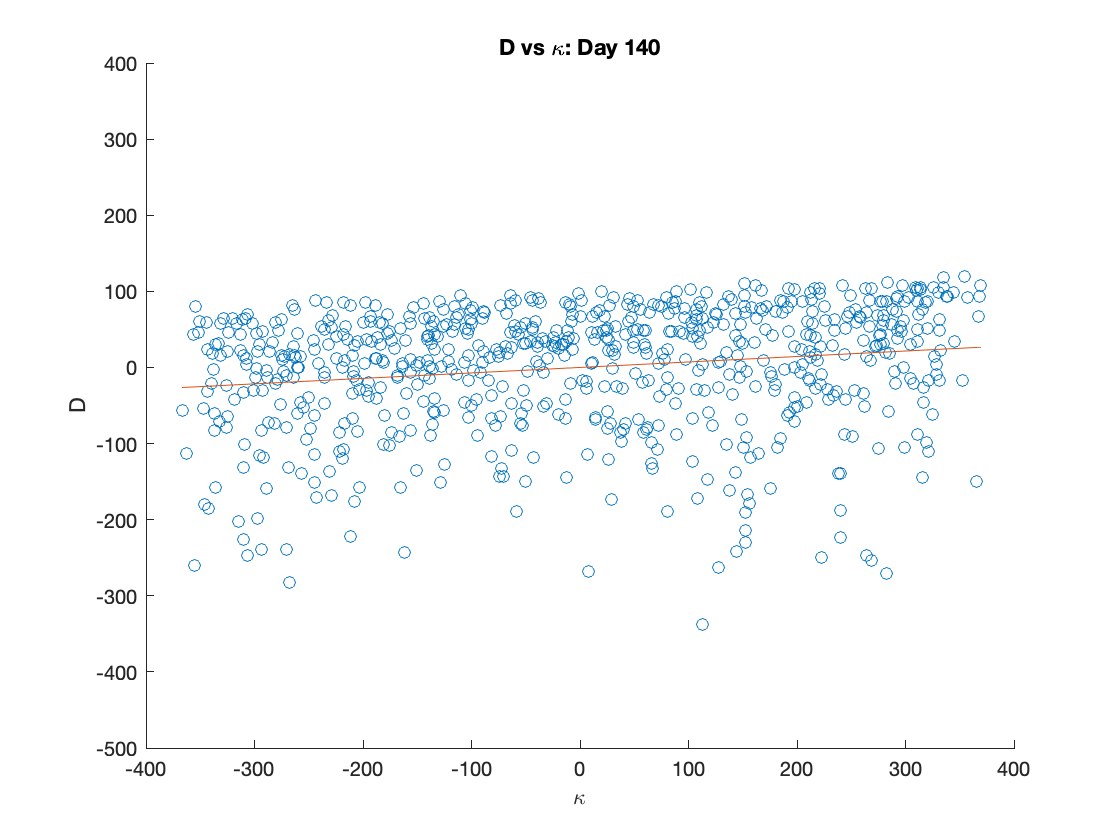}
		\caption{$D(t)$ vs $\kappa$: Day 140}
		\label{subfig:Dkappa140}
	\end{subfigure}
	\begin{subfigure}{0.475\textwidth}
		\includegraphics[width=\textwidth]{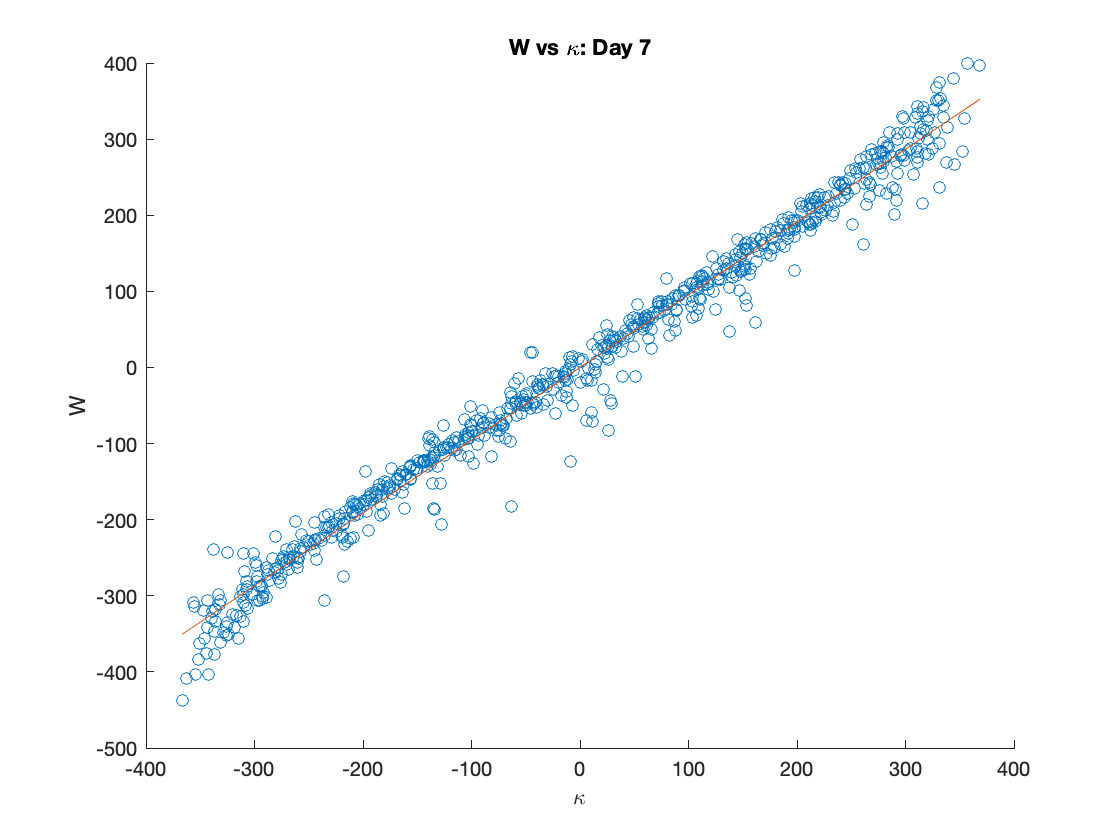}
		\caption{$W(t)$ vs $\kappa$: Day 7}
		\label{subfig:Wkappa7}
	\end{subfigure}
	\begin{subfigure}{0.475\textwidth}
		\includegraphics[width=\textwidth]{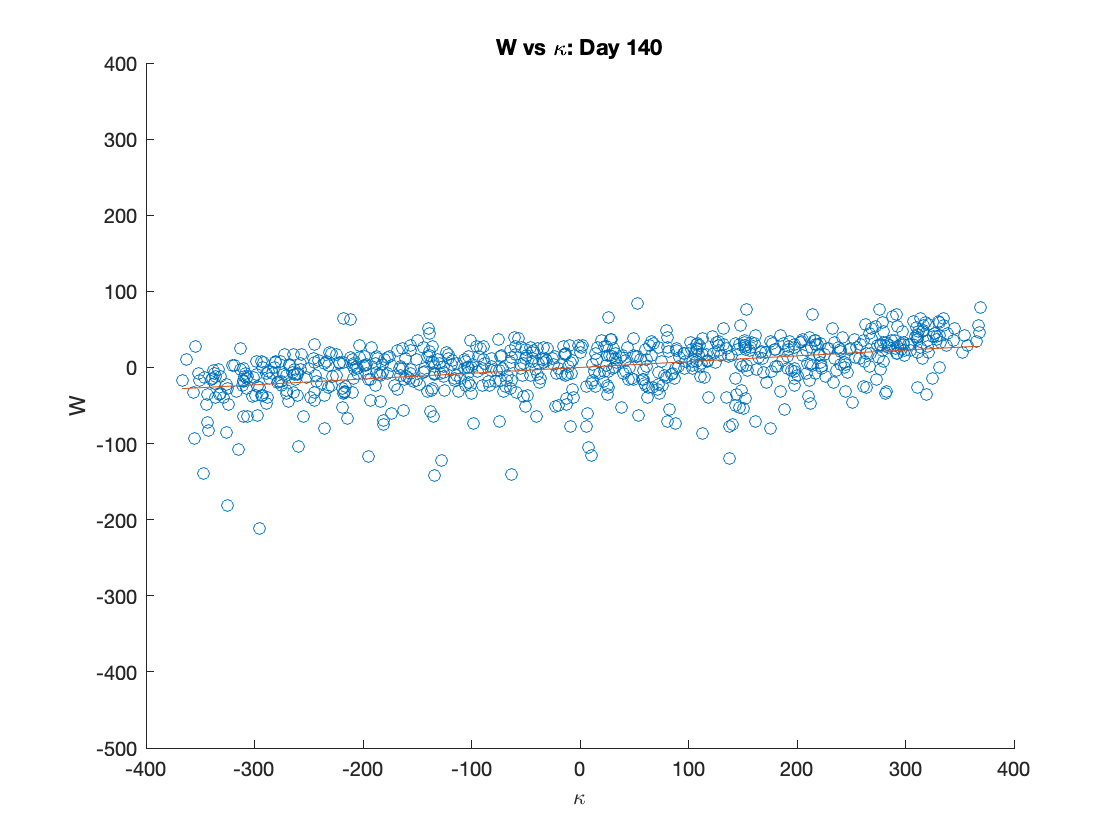}
		\caption{$W(t)$ vs $\kappa$: Day 140}
		\label{subfig:Wkappa140}
	\end{subfigure}
	\caption{Scatter plots of the residuals of the rankings for $D$ (Figure \ref{subfig:Dkappa7} and \ref{subfig:Dkappa140}) and $W$ (Figure \ref{subfig:Wkappa7} and \ref{subfig:Wkappa140}) with respect to the residuals of the $\kappa$ ranking.}
	\label{fig:kappascatter}
\end{figure}

When we look at the scatter plots of $W$ versus $\kappa$ in Figure \ref{fig:kappascatter}, we observe that the corresponding best-fit line at Day 7 is significantly steeper than the corresponding best-fit line at Day 140. We expect this result since $\kappa$ is the parameter that increases the count of infected individuals, but this increase slows down eventually (as shown in Figure \ref{fig:modelsim}). On the other hand, we see the same pattern in the scatter plots for $D$ versus $\kappa$ as the one observed in $D$ versus $\beta$: relatively flat best-fit lines and the points in the scatter plots are more dispersed towards the end of the simulation period. This suggests that the death count is only slightly sensitive to the reciprocal of the latency period.

\begin{figure}
	\centering
	\begin{subfigure}{0.475\textwidth}
		\includegraphics[width=\textwidth]{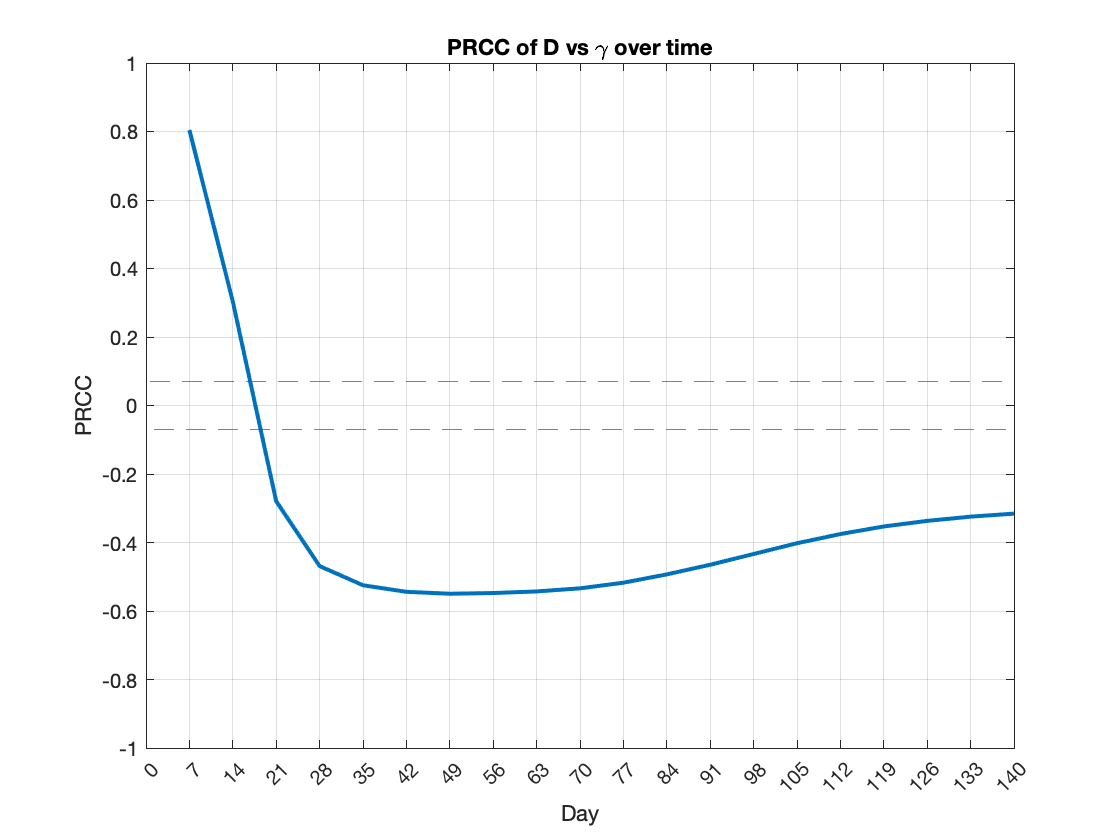}
		\caption{$D(t)$ vs $\gamma$}
		\label{subfig:Dgamma}
	\end{subfigure}
	\begin{subfigure}{0.475\textwidth}
		\includegraphics[width=\textwidth]{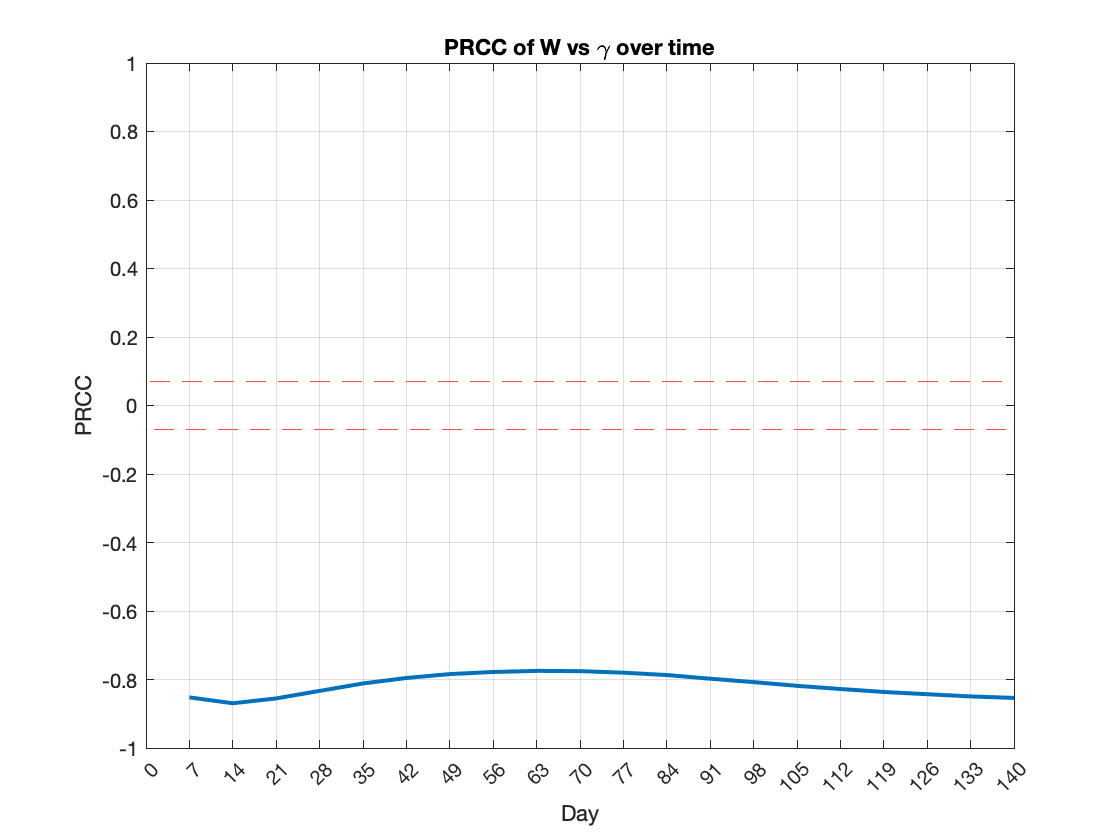}
		\caption{$W(t)$ vs $\gamma$}
		\label{subfig:Wgamma}
	\end{subfigure}
	\caption{Evolution of the PRCC of the output parameters $D$ (Figure \ref{subfig:Dgamma}) and $W$ (Figure \ref{subfig:Wgamma}) with respect to $\gamma$. The region between the two red lines indicate PRCC values that are not significantly different from 0 (where $p$-values are greater than 0.05), which is not the case for $\gamma$.}
	\label{fig:gamma}
\end{figure}

Figure \ref{fig:gamma} shows the evolution of the PRCC of the output variables $D$ and $W$ with respect to $\gamma$. We observe a strong negative correlation of $W$ with $\gamma$, which is to be expected as $\gamma$ is the parameter involved in decreasing the infected compartments $I$ and $I_v$. However, we observe that $D$ initially shows a positive correlation with $\gamma$ (up to Day 14), and then we eventually observe a negative correlation. The corresponding $p$-values are very small, indicating that the PRCC values are significantly different from 0.

\begin{figure}
	\centering
	\begin{subfigure}{0.475\textwidth}
		\includegraphics[width=\textwidth]{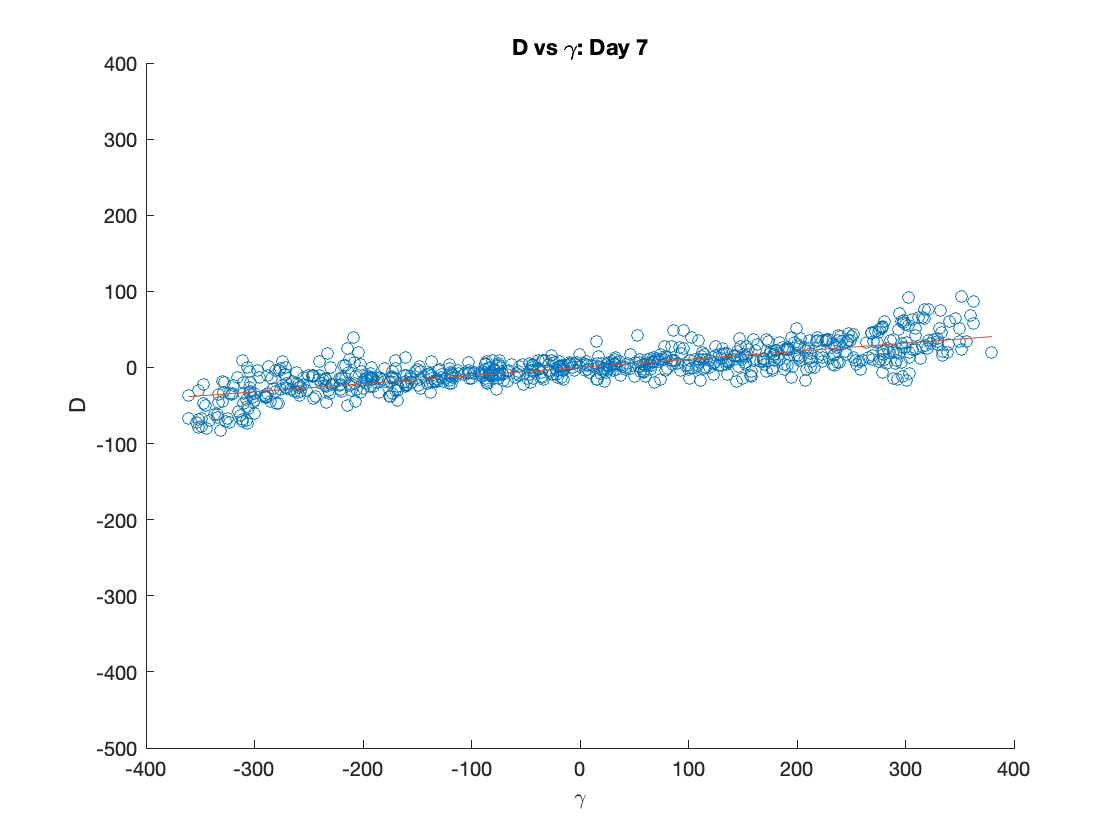}
		\caption{$D(t)$ vs $\gamma$: Day 7}
		\label{subfig:Dgamma7}
	\end{subfigure}
	\begin{subfigure}{0.475\textwidth}
		\includegraphics[width=\textwidth]{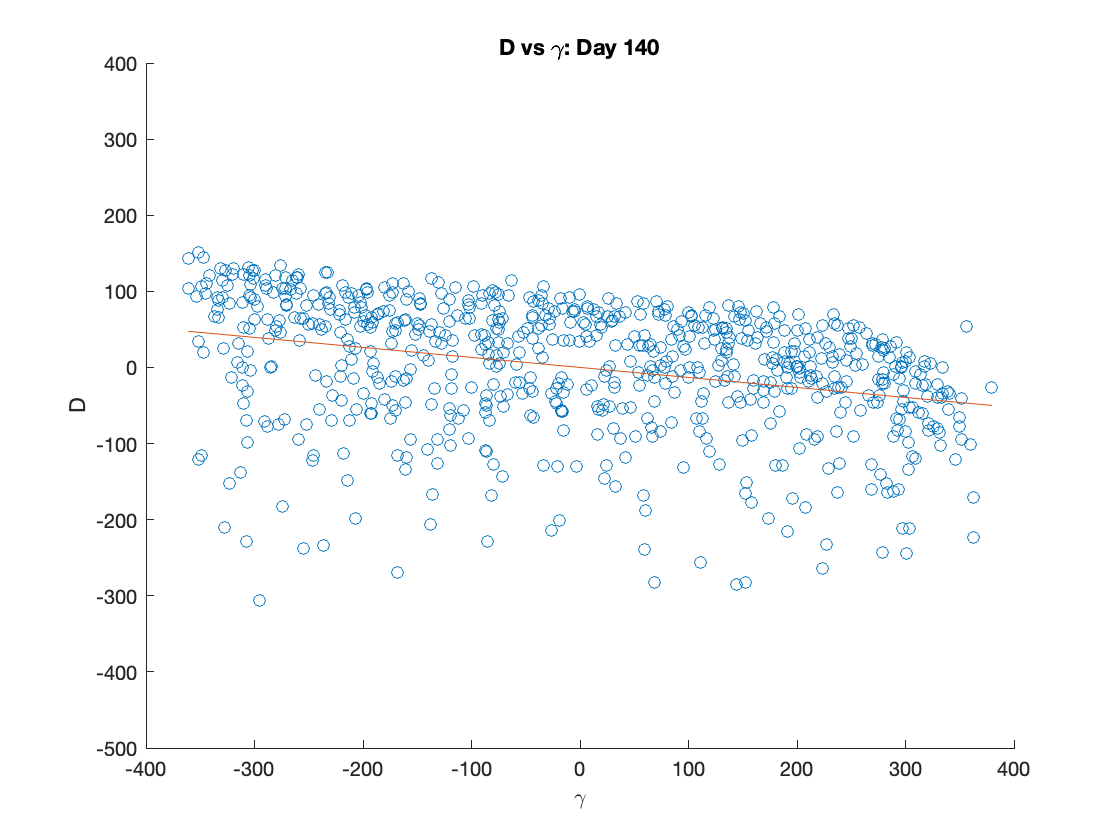}
		\caption{$D(t)$ vs $\gamma$: Day 140}
		\label{subfig:Dgamma140}
	\end{subfigure}
	\begin{subfigure}{0.475\textwidth}
		\includegraphics[width=\textwidth]{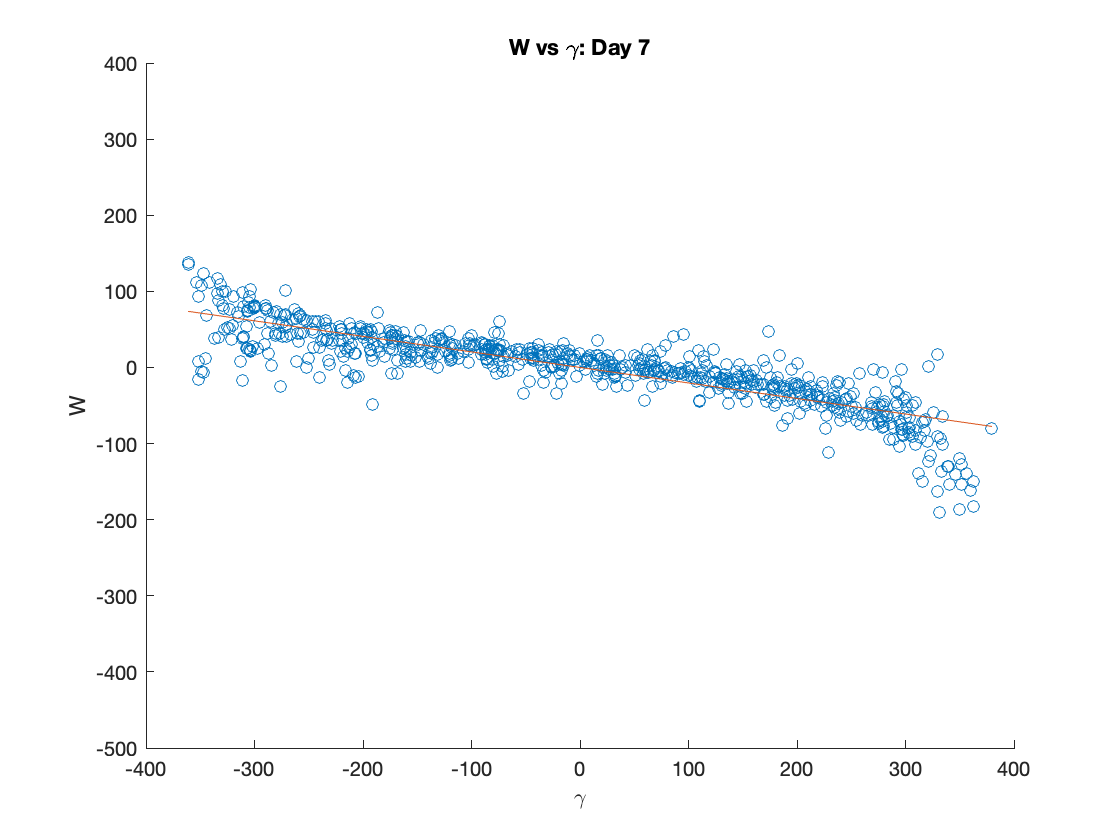}
		\caption{$W(t)$ vs $\gamma$: Day 7}
		\label{subfig:Wgamma7}
	\end{subfigure}
	\begin{subfigure}{0.475\textwidth}
		\includegraphics[width=\textwidth]{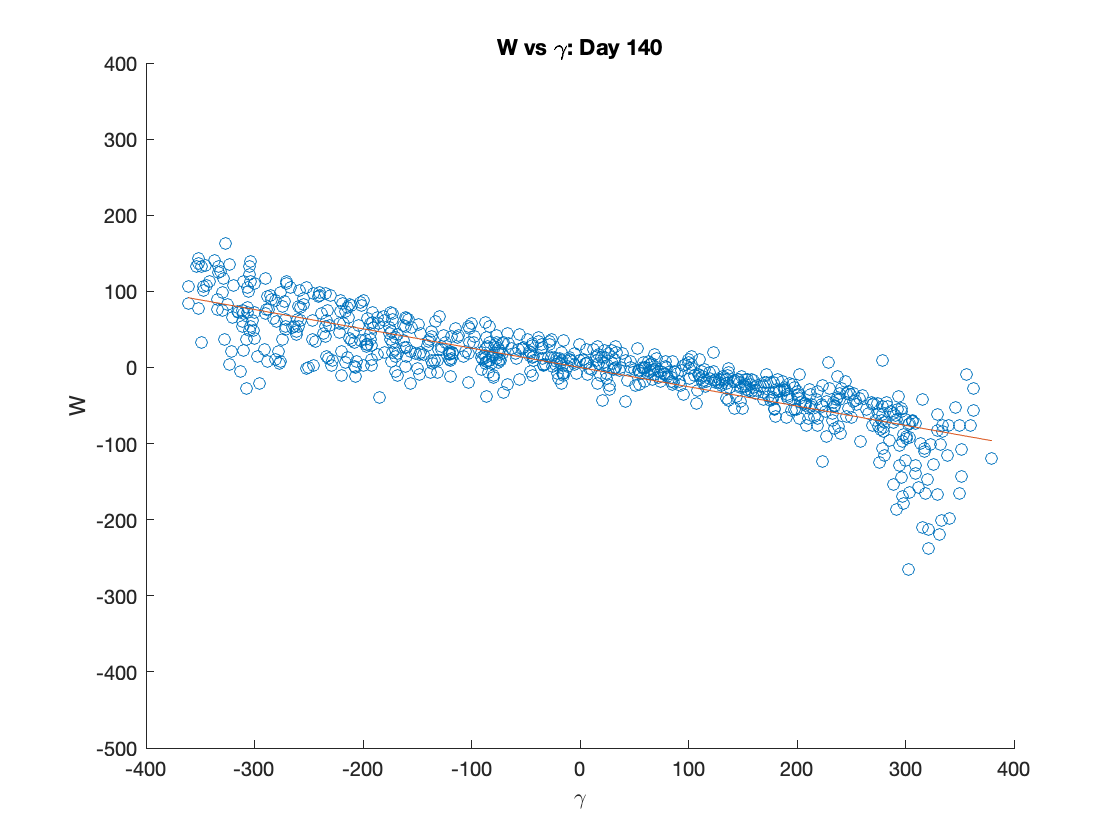}
		\caption{$W(t)$ vs $\gamma$: Day 140}
		\label{subfig:Wgamma140}
	\end{subfigure}
	\caption{Scatter plots of the residuals of the rankings for $D$ (Figure \ref{subfig:Dgamma7} and \ref{subfig:Dgamma140}) and $W$ (Figure \ref{subfig:Wgamma7} and \ref{subfig:Wgamma140}) with respect to the residuals of the $\gamma$ ranking.}
	\label{fig:gammascatter}
\end{figure}

Figure \ref{fig:gammascatter} shows the scatter plots of the output variables with respect to $\gamma$. Notice that the slope of the best-fit line for $D$ is positive at Day 7 but negative at Day 140. We also observe that the points are closer to the best-fit line at Day 7 and become more dispersed at Day 140. In contrast, the negative slope of the best-fit line for $W$ is sustained throughout the simulation period, with the points located close to the best-fit line. For both cases, the best-fit lines are not so steep, implying that the output variables are only slightly sensitive to $\gamma$.

\begin{figure}
	\centering
    \includegraphics[width=0.75\textwidth]{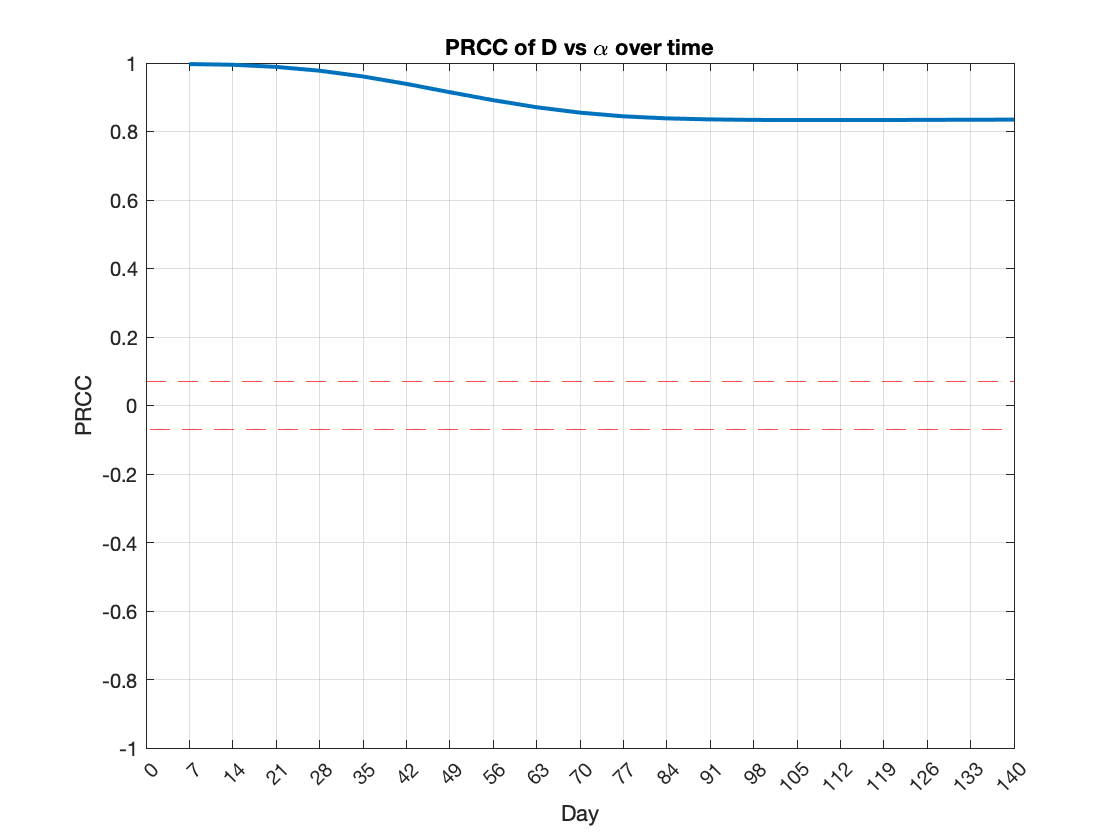}
	\caption{Evolution of the PRCC of output parameter $D$ with respect to $\alpha$. The region between the two red lines indicate PRCC values that are not significantly different from 0 (where $p$-values are greater than 0.05), which is not the case for $\alpha$.}
	\label{fig:Dalpha}
\end{figure}

Due to the non-monotonic relationship of $W$ with $\alpha$, we cannot use LHS-PRCC for sensitivity analysis. Because of this we shall only look at the evolution of the PRCC values of $D$ with respect to $\alpha$ (unlike the previous discussions with the other disease parameters), as shown in Figure \ref{fig:Dalpha}. We observe a very strong positive correlation that only slightly decreases over time. The $p$-values are close to 0, implying that the PRCC values are significantly different from 0.

\begin{figure}
	\centering
	\begin{subfigure}{0.475\textwidth}
		\includegraphics[width=\textwidth]{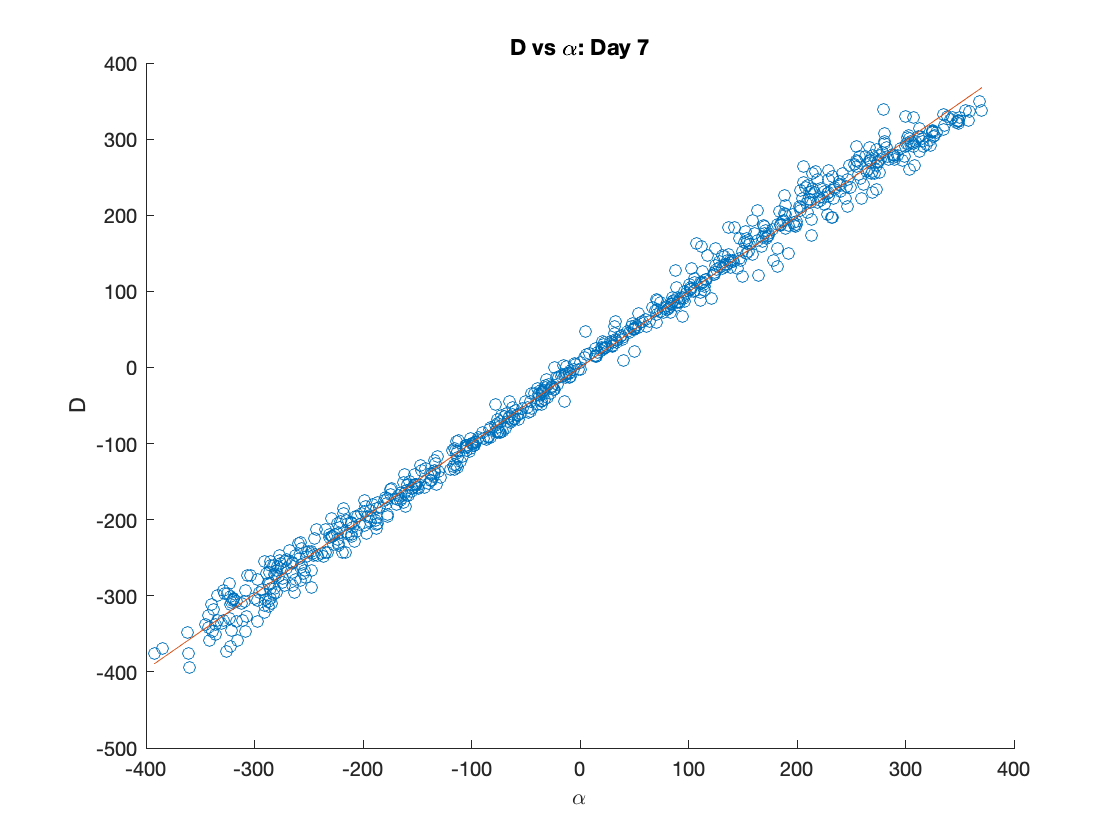}
		\caption{$D(t)$ vs $\alpha$: Day 7}
		\label{subfig:Dalpha7}
	\end{subfigure}
	\begin{subfigure}{0.475\textwidth}
		\includegraphics[width=\textwidth]{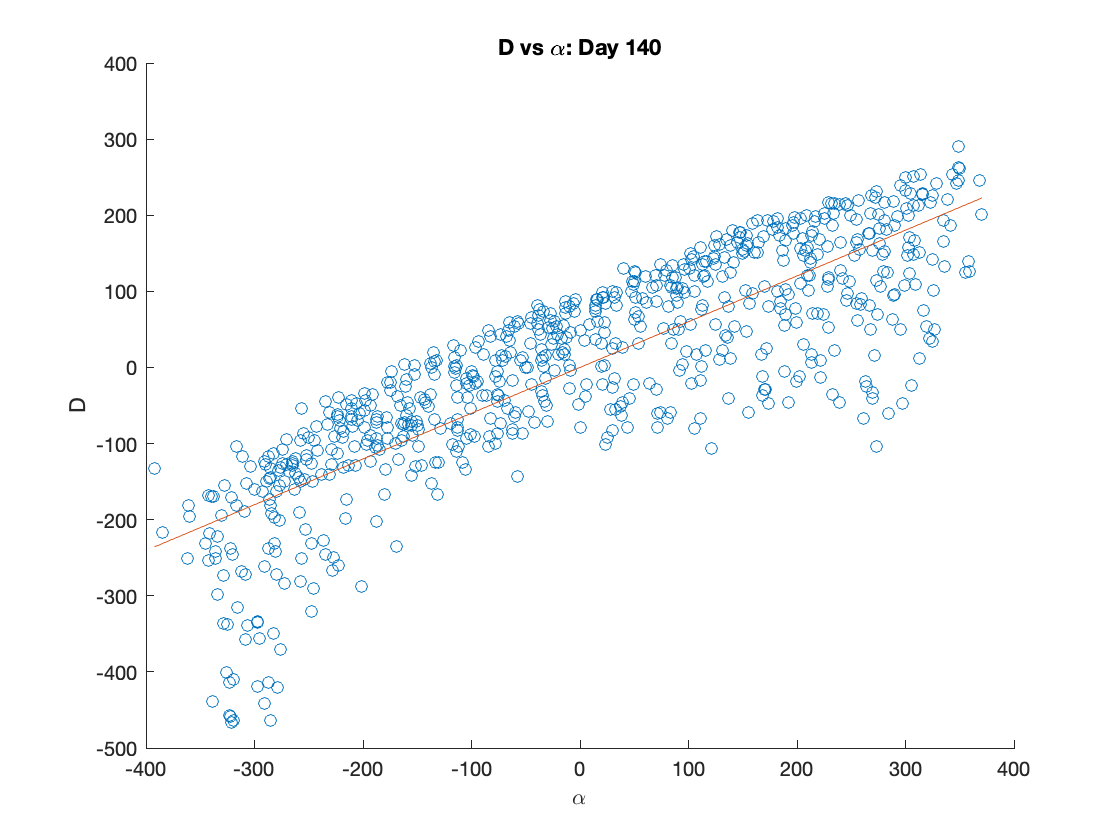}
		\caption{$D(t)$ vs $\alpha$: Day 140}
		\label{subfig:Dalpha21}
	\end{subfigure}
	\caption{Scatter plots of the residuals of the rankings for $D$ with respect to the residuals of the $\alpha$ ranking.}
	\label{fig:Dalphascatter}
\end{figure}

We present the scatter plots of the residuals of $D$ versus the residuals of $\alpha$ in Figure \ref{fig:Dalphascatter}. Notice that the slope of the best-fit line decreases only slightly towards the end of the simulation period. However, the dispersion of the points near the best-fit line are higher at Day 140 compared with Day 7. These still show the high sensitivity of $D$ with $\alpha$, indicating that efforts to reduce the death ratio (by finding a cure to the disease or simply improving the overall healthcare system) are worthwhile.

To summarize, the cumulative death count $D$ is sensitive to all disease parameters, but the sensitivity is not that strong, save for the death ratio $\alpha$. This highlights the importance of reducing $\alpha$ in controlling the death count, either by finding a cure for the disease or improving the healthcare system. On the other hand, the cumulative infected count $W$ is not very sensitive to the disease transmission rate $\beta$ and the reciprocal of the infectious period $\gamma$. However, $W$ shows a strong sensitivity to the reciprocal of the latency period ($\kappa$) at the start of the simulation, which gradually decreases over time.

\subsection{Sensitivity with Respect to Vaccine Parameters}
\begin{figure}
	\centering
	\begin{subfigure}{0.475\textwidth}
		\includegraphics[width=\textwidth]{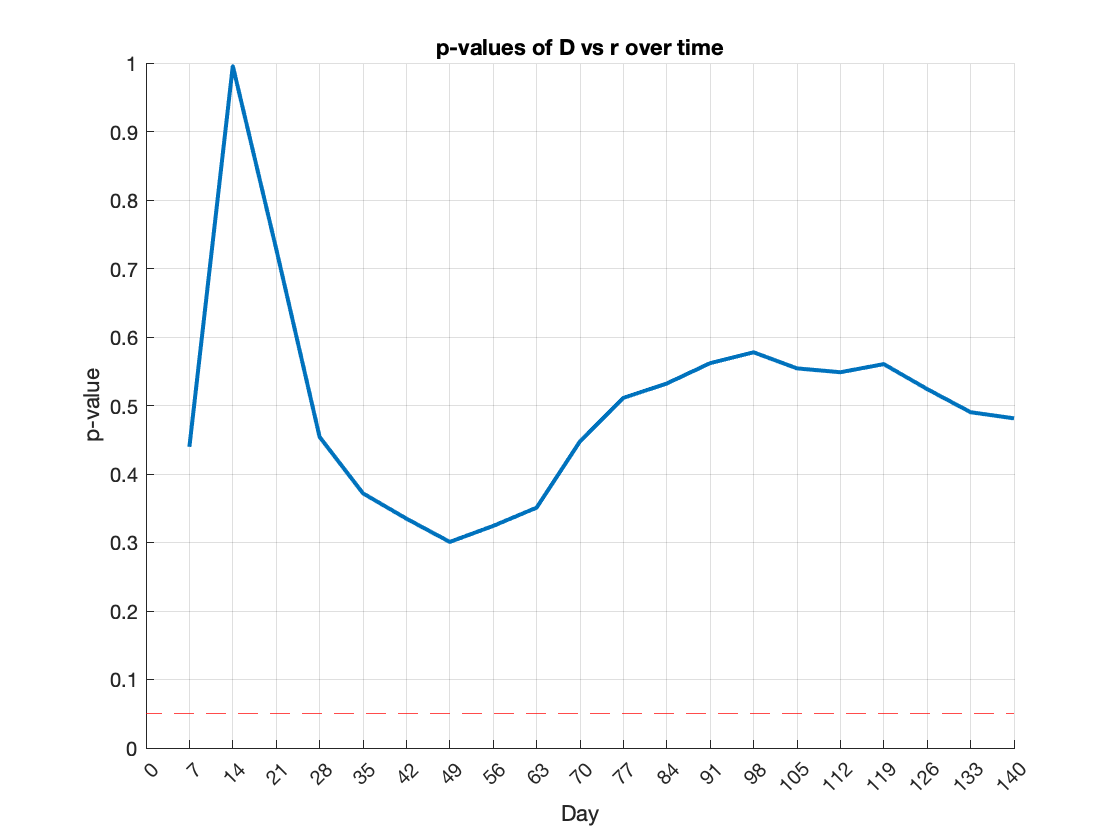}
		\caption{$p$-values of $D(t)$ vs $r$}
		\label{subfig:Drpval}
	\end{subfigure}
	\begin{subfigure}{0.475\textwidth}
		\includegraphics[width=\textwidth]{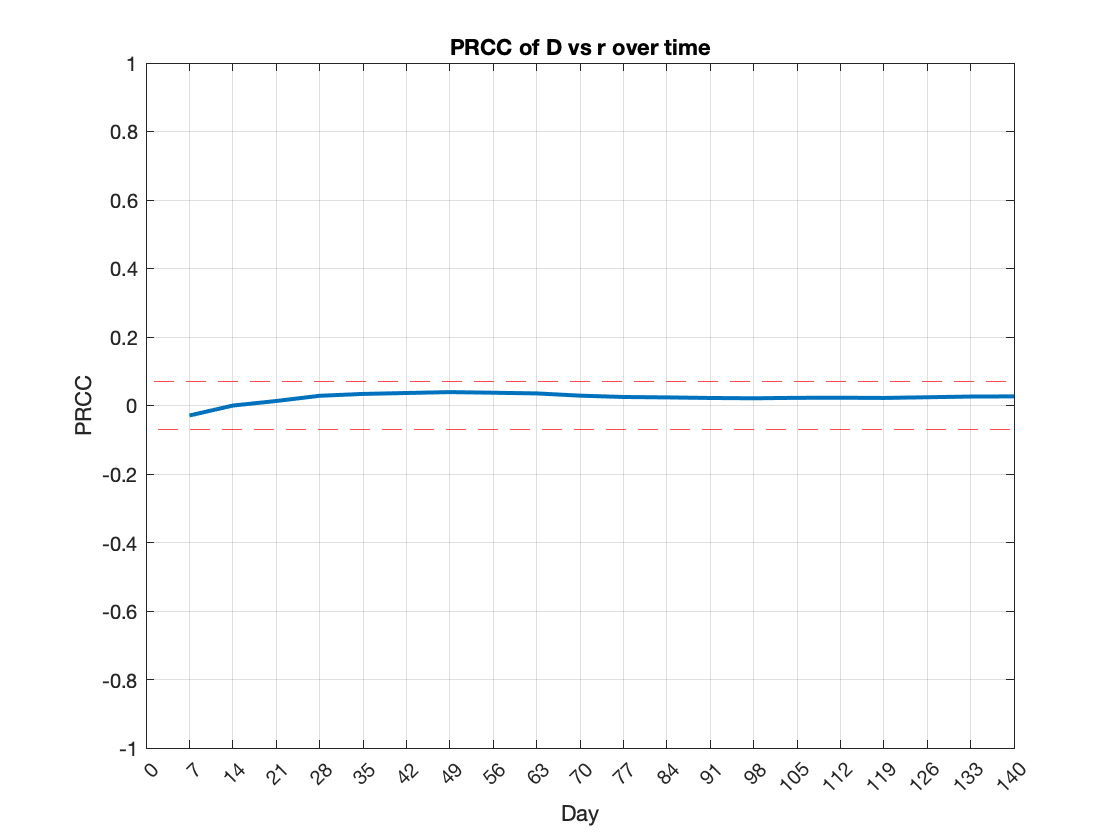}
		\caption{PRCC of $D(t)$ vs $r$}
		\label{subfig:Drprcc}
	\end{subfigure}
	\caption{Evolution of the $p$-values (Figure \ref{subfig:Drpval}) and the PRCC (Figure \ref{subfig:Drprcc}) of the output parameter $D$ with respect to $r$. The red lines indicate the threshold separating PRCC values that are not significantly different from 0. These plots show that the PRCC values are not statistically different from 0 on the entire simulation period.}
	\label{fig:Dr}
\end{figure}

Figure \ref{fig:Dr} shows the evolution of $p$-values and PRCC values for the output variable $D$ with respect to $r$. We observe that the $p$-values are consistently above the threshold level of significance of 0.05 (Figure \ref{subfig:Drpval}), indicating that there is not enough information to say that the PRCC values are different from 0 as verified by Figure \ref{subfig:Drprcc}. This may imply that $D$ is not sensitive to $r$. It is worth noting that there is no apparent effect of the vaccine efficacy on the cumulative death count, even though we expect the distribution of the vaccine to reduce the number of infected individuals. This may be due to the imperfect nature of the vaccine, which does not prevent the individual from being infected.

\begin{figure}
	\centering
	\begin{subfigure}{0.475\textwidth}
		\includegraphics[width=\textwidth]{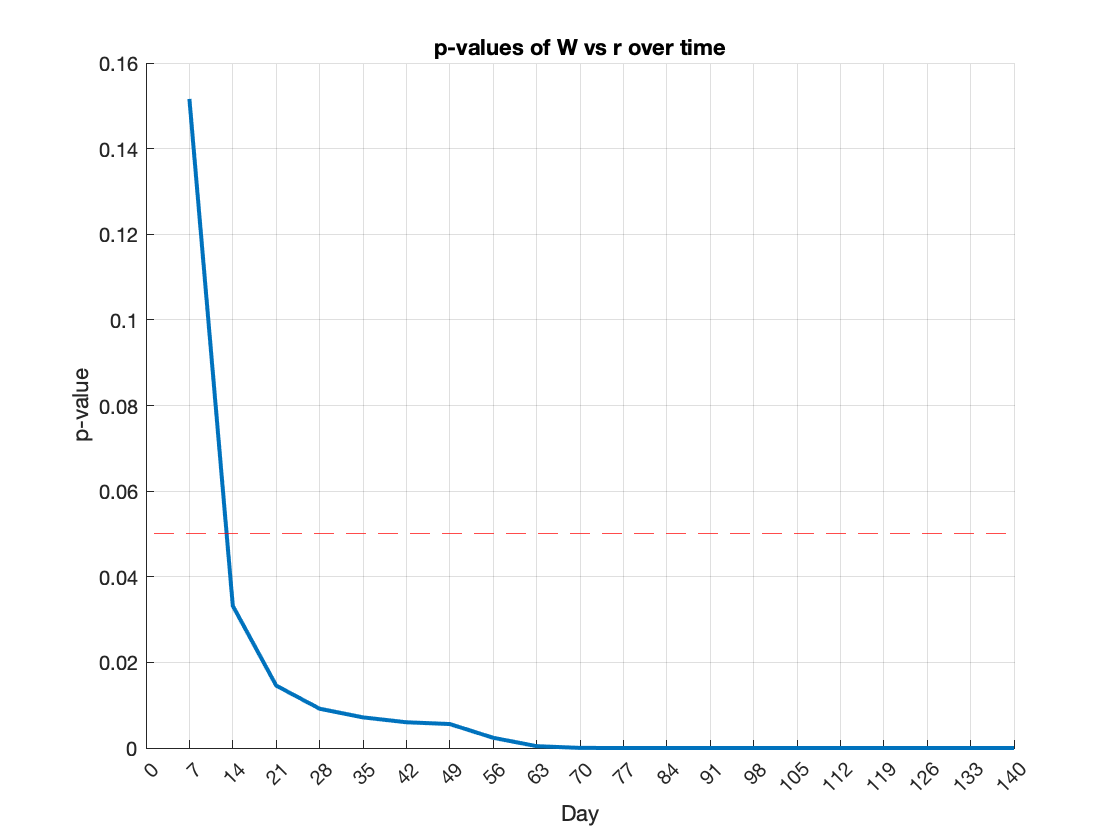}
		\caption{$p$-values of $W(t)$ vs $r$}
		\label{subfig:Wrpval}
	\end{subfigure}
	\begin{subfigure}{0.475\textwidth}
		\includegraphics[width=\textwidth]{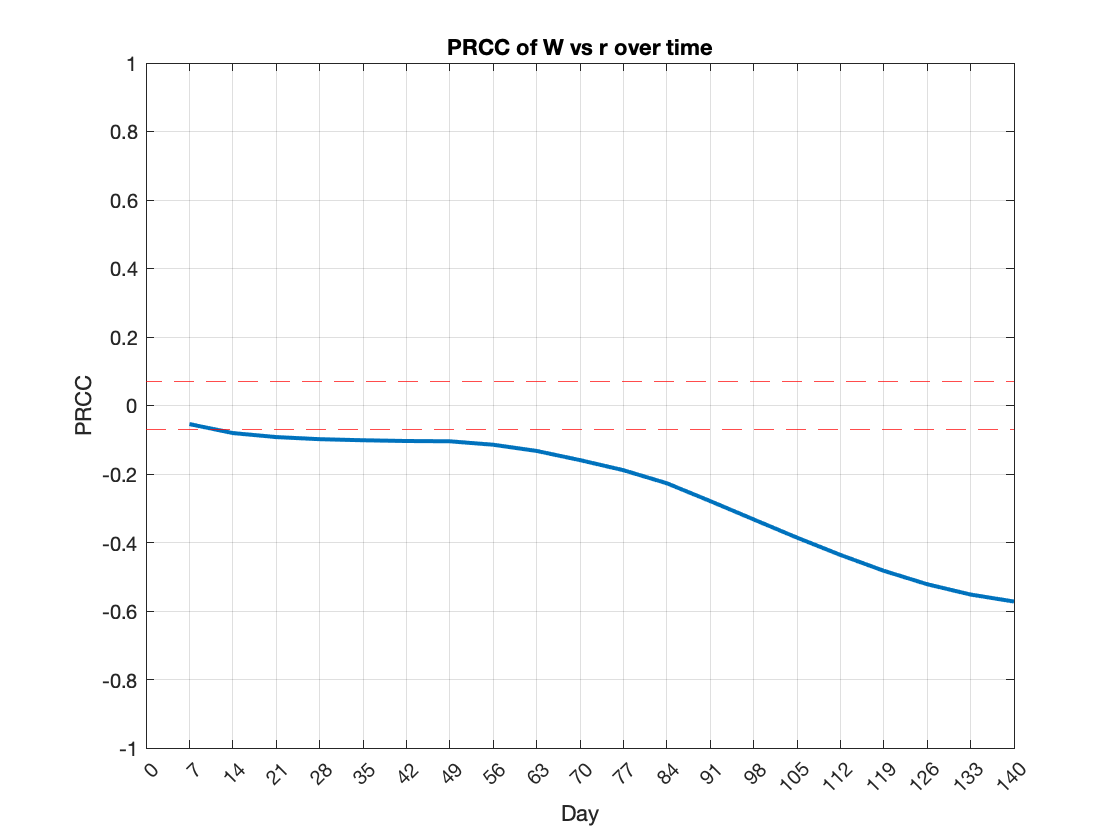}
		\caption{PRCC of $W(t)$ vs $r$}
		\label{subfig:Wrprcc}
	\end{subfigure}
	\caption{Evolution of the $p$-values (Figure \ref{subfig:Wrpval}) and the PRCC (Figure \ref{subfig:Wrprcc}) of the output variable $W$ with respect to $r$. The red lines indicate the threshold separating PRCC values that are not significantly different from 0. These plots show that the PRCC values are statistically different from 0 starting from Day 14.}
	\label{fig:Wr}
\end{figure}

Figure \ref{fig:Wr} shows the evolution of $p$-values and PRCC for the output variable $W$ with respect to $r$. We observe that the $p$-value is above the 0.05 threshold on Day 7 but falls below the threshold starting from Day 14. This implies that the PRCC values are significantly different from 0 starting from Day 14, indicating the delayed effect of the vaccine efficacy on the cumulative count of infected individuals.

\begin{figure}
	\centering
	\begin{subfigure}{0.475\textwidth}
		\includegraphics[width=\textwidth]{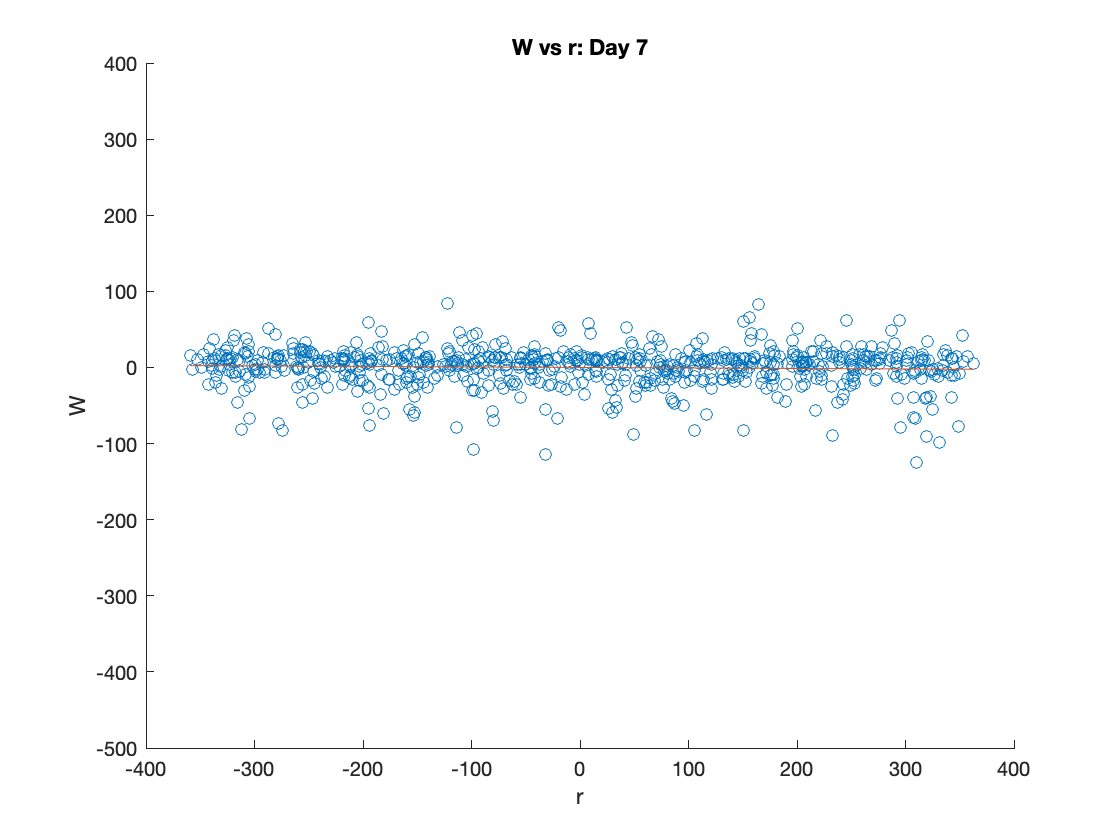}
		\caption{$W(t)$ vs $r$: Day 7}
		\label{subfig:Wr28}
	\end{subfigure}
	\begin{subfigure}{0.475\textwidth}
		\includegraphics[width=\textwidth]{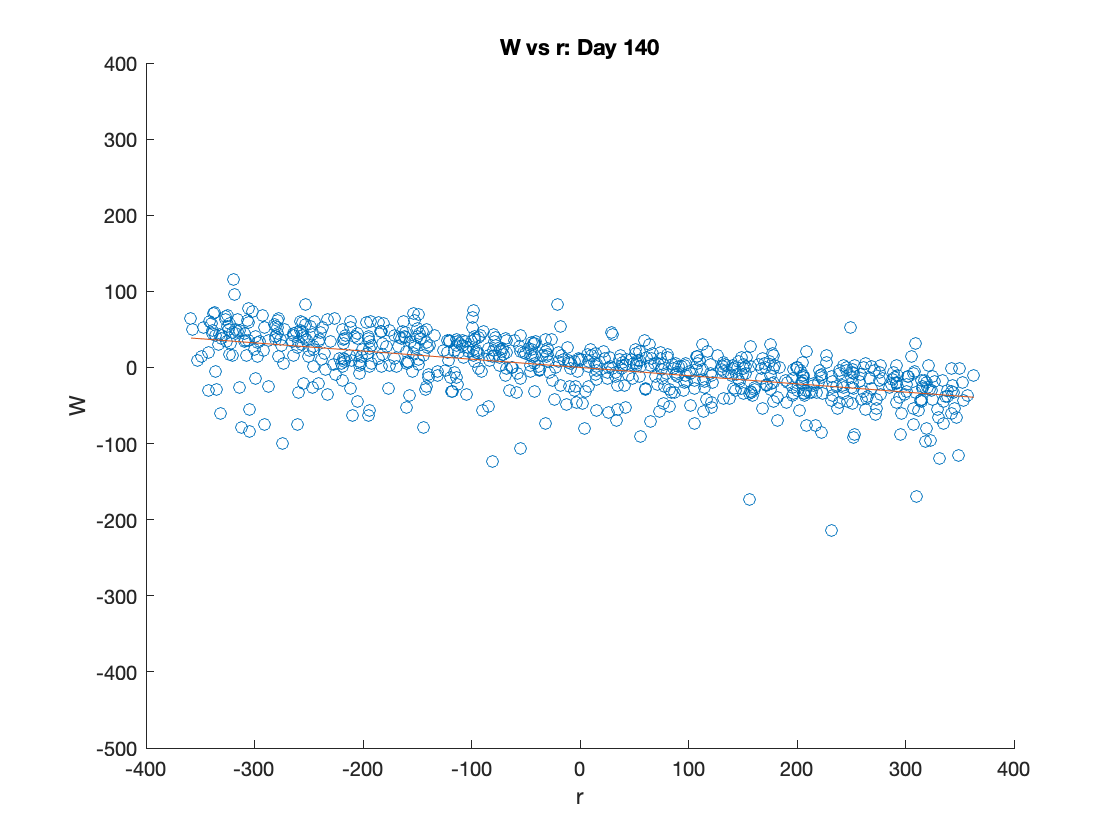}
		\caption{$W(t)$ vs $r$: Day 140}
		\label{subfig:Wr140}
	\end{subfigure}
	\caption{Scatter plots of the residuals of the rankings for $W$ with respect to the residuals of the $r$ ranking.}
	\label{fig:Wrscatter}
\end{figure}

Figure \ref{fig:Wrscatter} shows the scatter plots of the resuduals of $W$ versus the residuals of $r$. Notice that the best-fit line is almost horizontal on Day 7, but becomes decreasing at Day 140. The best-fit line is still relatively horizontal, implying that $W$ is only slightly sensitive to $r$. It may not be worthwhile to wait for a vaccine with a higher efficacy rate.

\begin{figure}
	\centering
	\begin{subfigure}{0.475\textwidth}
		\includegraphics[width=\textwidth]{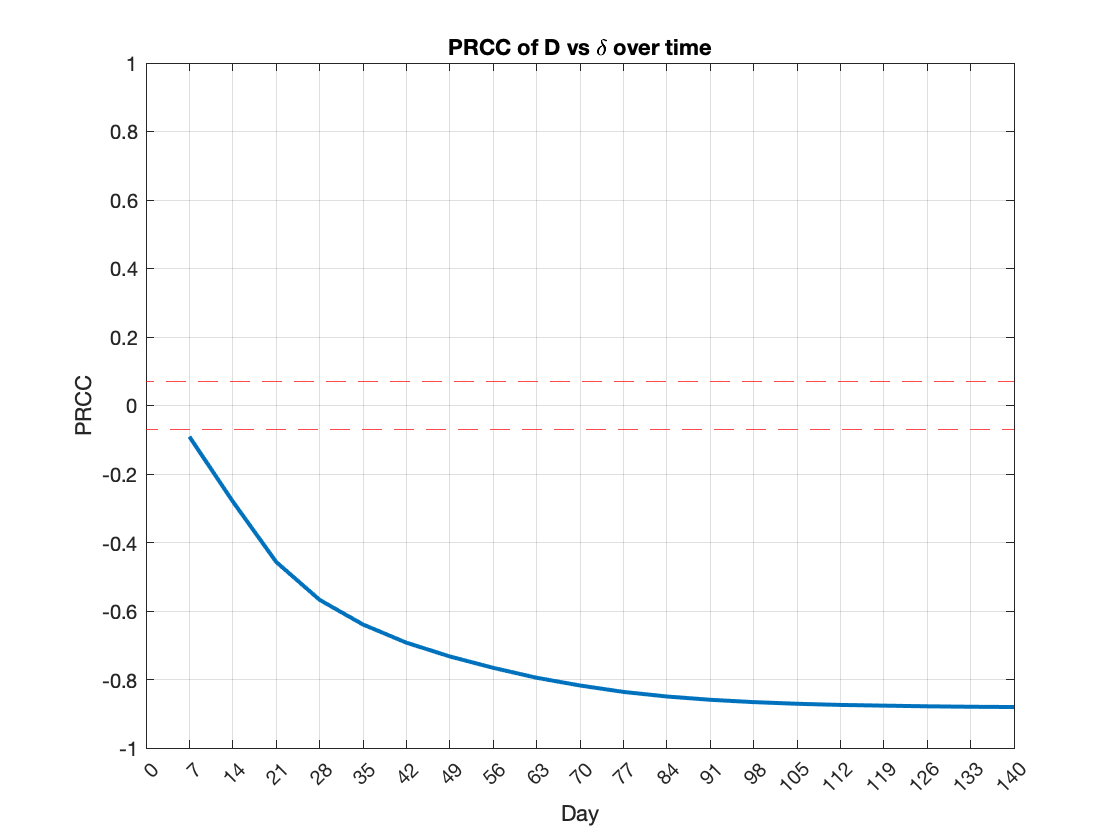}
		\caption{$D(t)$ vs $\delta$}
		\label{subfig:Ddelta}
	\end{subfigure}
	\begin{subfigure}{0.475\textwidth}
		\includegraphics[width=\textwidth]{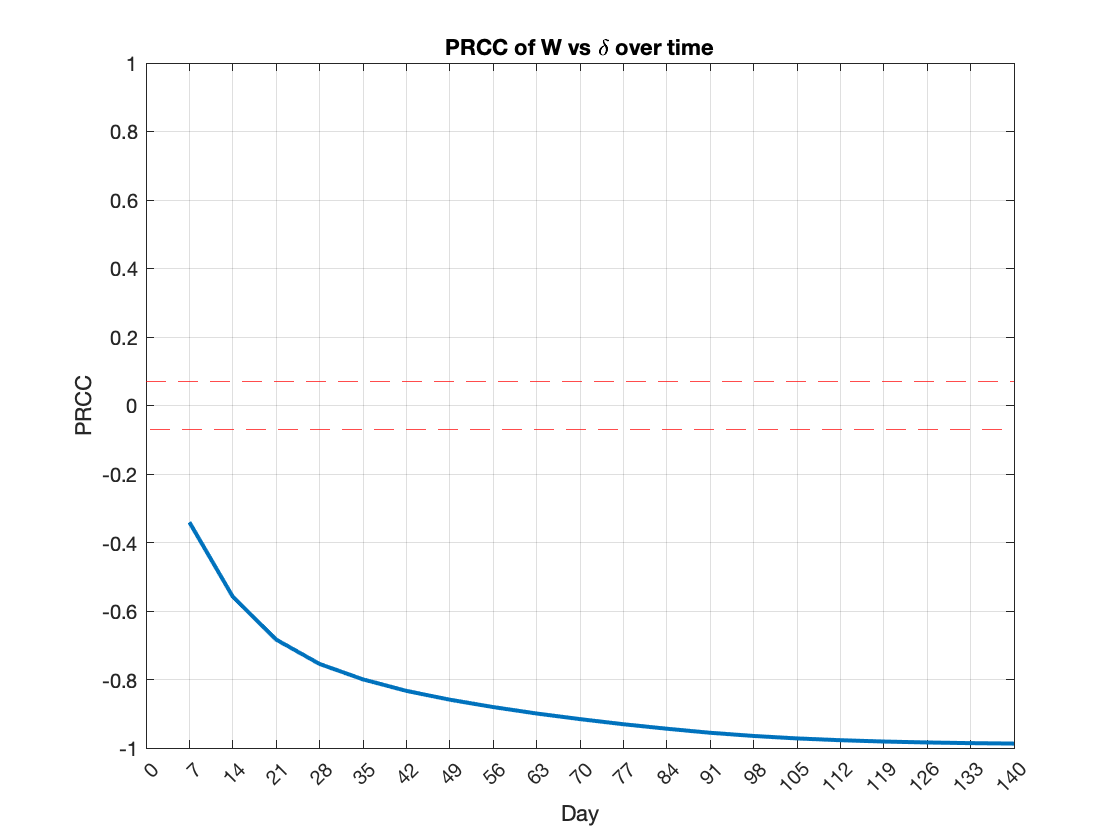}
		\caption{$W(t)$ vs $\delta$}
		\label{subfig:Wdelta}
	\end{subfigure}
	\caption{Evolution of the PRCC of the output parameters $D$ (Figure \ref{subfig:Ddelta}) and $W$ (Figure \ref{subfig:Wdelta}) with respect to $\delta$. The region between the two red lines indicate PRCC values that are not significantly different from 0 (where $p$-values are greater than 0.05), which is not the case for $\delta$.}
	\label{fig:delta}
\end{figure}

Figure \ref{fig:delta} shows the evolution of the PRCC values of the output variables $D$ and $W$ with respect to $\delta$. We observe that both output variables are negatively correlated to $\delta$, which is to be expected since increasing the number of vaccinated individuals will lead to lower chances of being infected and also dying from the disease. However, note that the correlation is relatively weak at the beginning of the simulation period and grows stronger as time passes. The corresponding $p$-values are all below the 0.05 threshold, indicating that the PRCC values are significanly different from 0.

\begin{figure}
	\centering
	\begin{subfigure}{0.475\textwidth}
		\includegraphics[width=\textwidth]{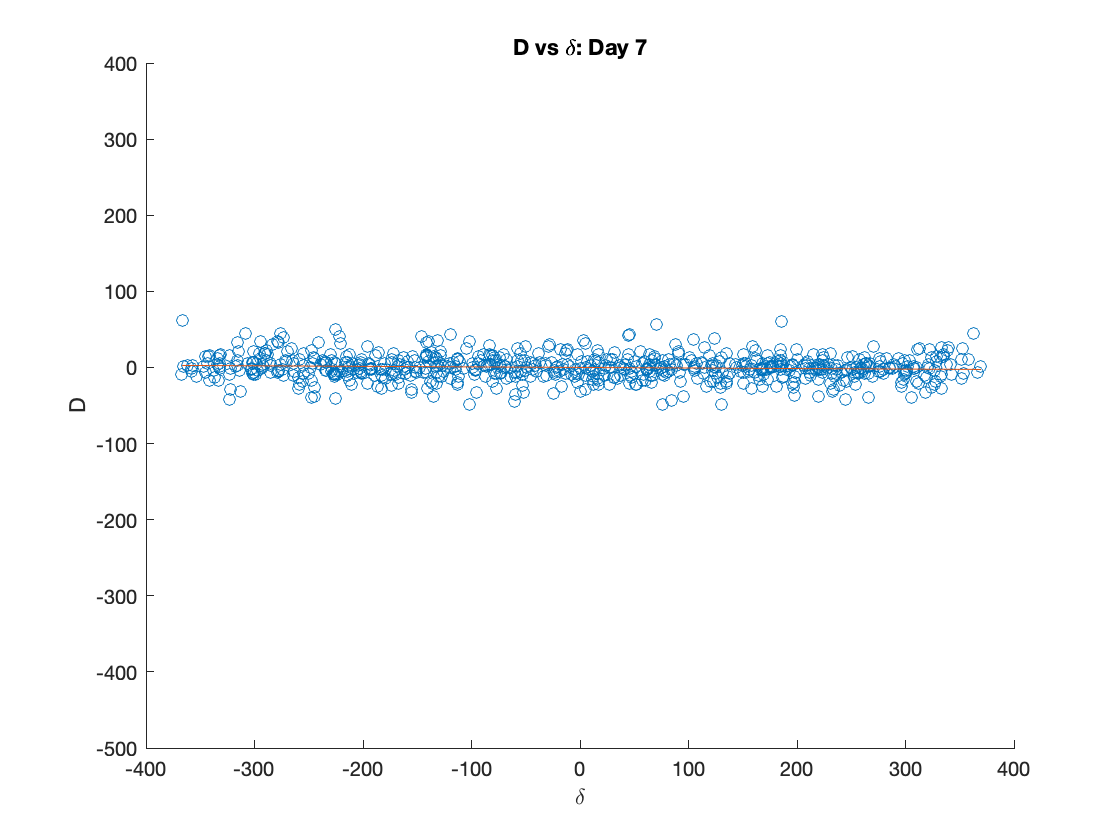}
		\caption{$D(t)$ vs $\delta$: Day 7}
		\label{subfig:Ddelta7}
	\end{subfigure}
	\begin{subfigure}{0.475\textwidth}
		\includegraphics[width=\textwidth]{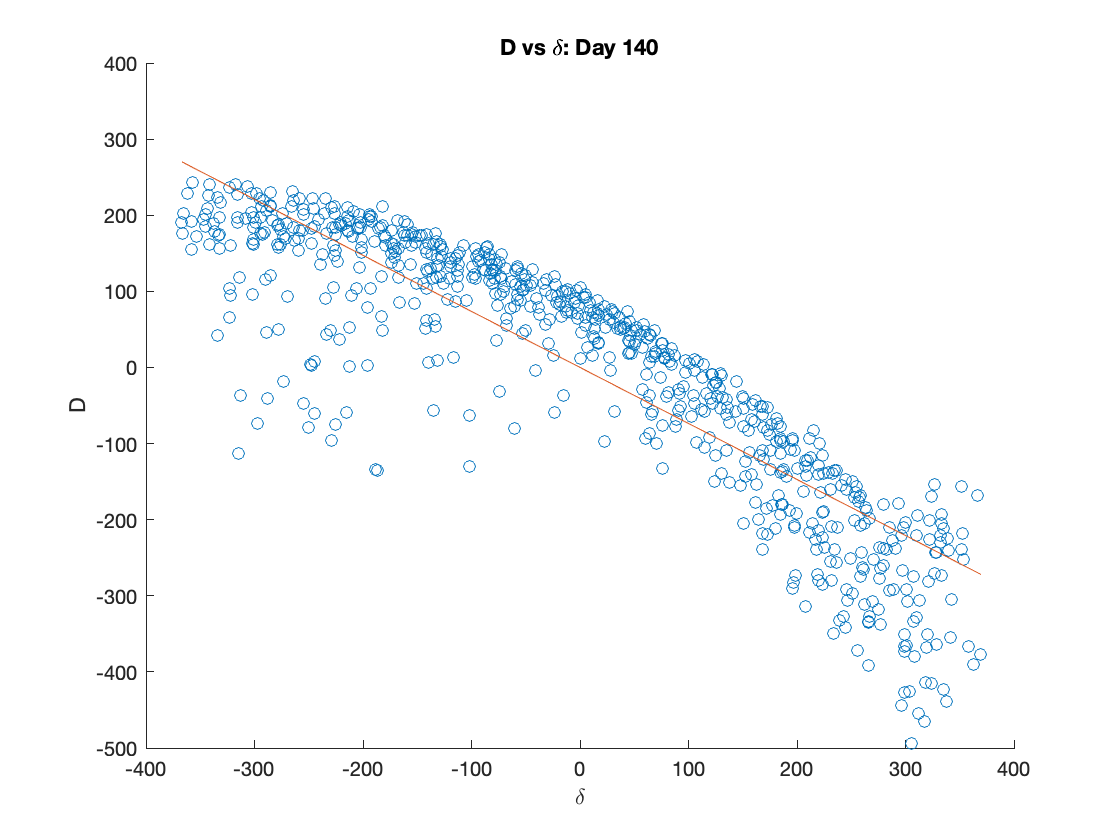}
		\caption{$D(t)$ vs $\delta$: Day 140}
		\label{subfig:Ddelta140}
	\end{subfigure}
	\begin{subfigure}{0.475\textwidth}
		\includegraphics[width=\textwidth]{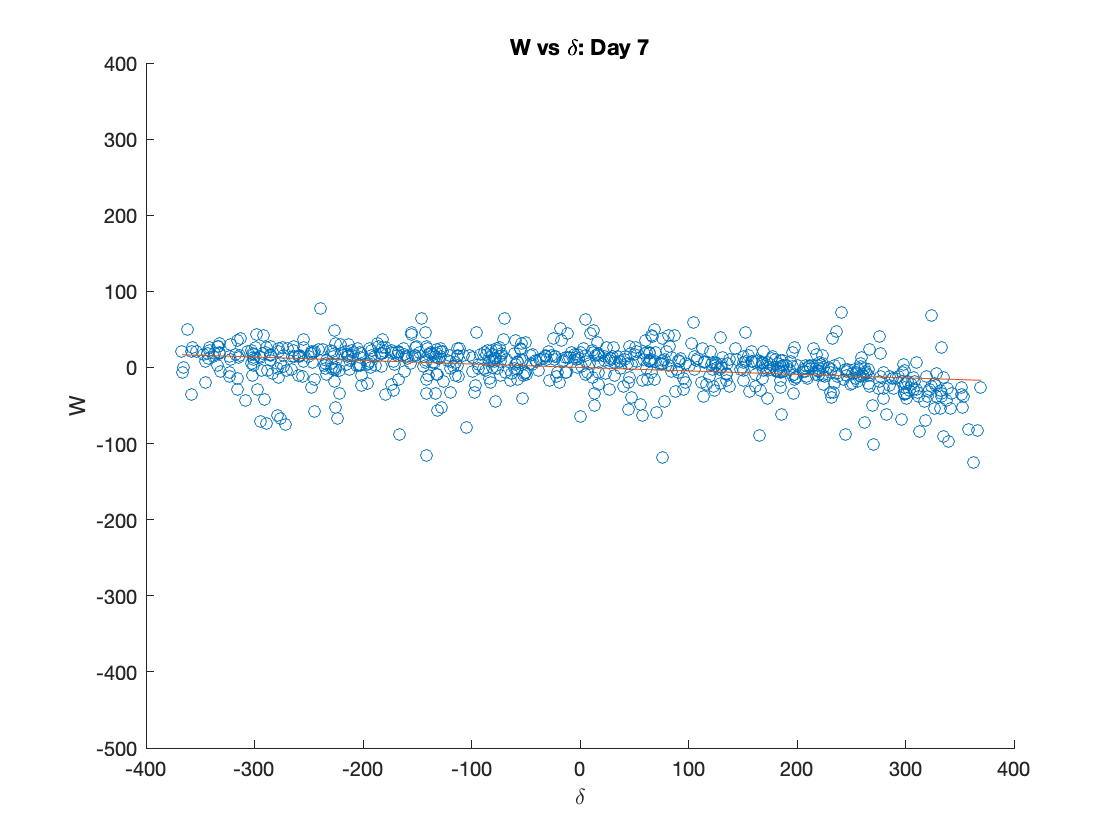}
		\caption{$W(t)$ vs $\delta$: Day 7}
		\label{subfig:Wdelta7}
	\end{subfigure}
	\begin{subfigure}{0.475\textwidth}
		\includegraphics[width=\textwidth]{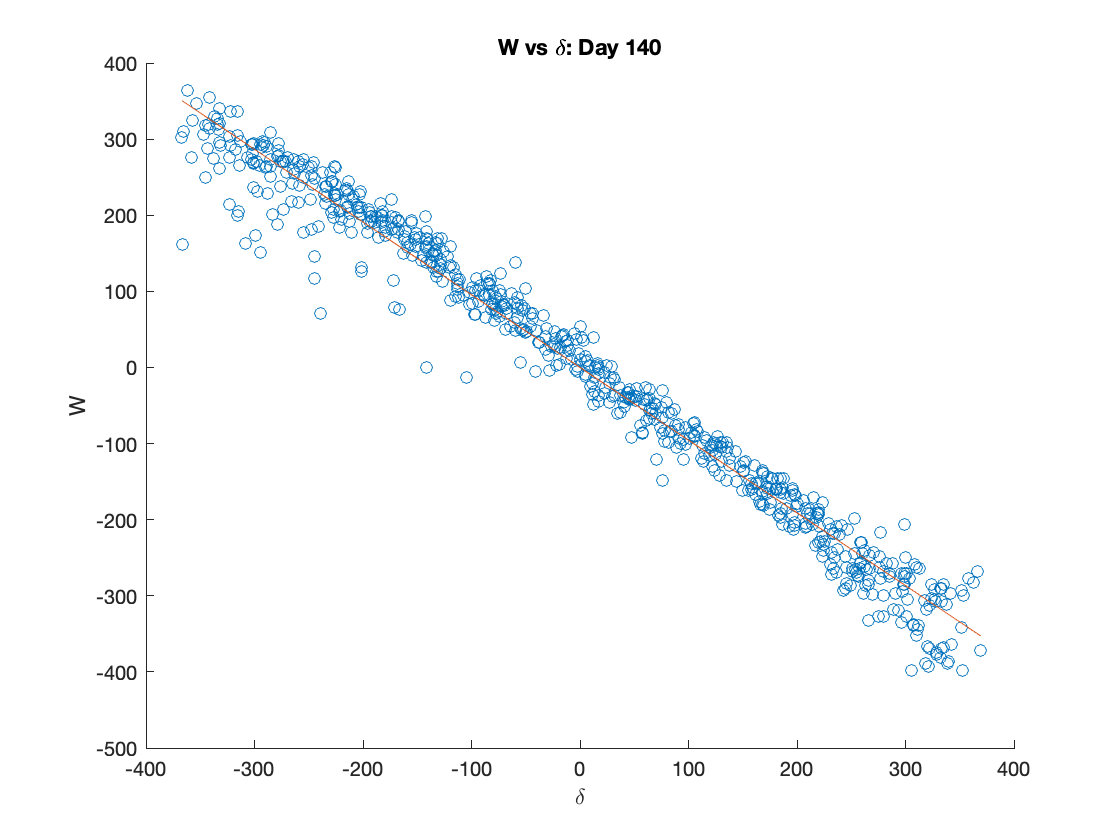}
		\caption{$W(t)$ vs $\delta$: Day 140}
		\label{subfig:Wdelta140}
	\end{subfigure}
	\caption{Scatter plots of the residuals of the rankings for $D$ (Figure \ref{subfig:Ddelta7} and \ref{subfig:Ddelta140}) and $W$ (Figure \ref{subfig:Wdelta7} and \ref{subfig:Wdelta140}) with respect to the residuals of the $\delta$ ranking.}
	\label{fig:deltascatter}
\end{figure}

The scatter plots for the residuals of the output variables versus the residuals of $\delta$ are shown in Figure \ref{fig:deltascatter}. At Day 7, the best-fit lines for both $D$ and $W$ are almost horizontal, but the decreasing relationship is evident towards the end of the simulation period (Day 140). It is also interesting to note that the points in the scatter plot of $D$ versus $\delta$ appear to form a curved line instead of a straight line. This may imply a nonlinear relationship between these two variables.

In summary, the cumulative death count and the cumulative infected count are both sensitive to the vaccine rollout rate, with the sensitivity growing stronger as time passes. On the other hand, only the cumulative infected count is sensitive to the vaccine efficacy rate, which is only evident after 2 weeks of vaccine distribution. Moreover, this sensitivity is not as strong compared with the vaccine rollout rate. This implies that the vaccine rollout rate is more important in curbing the spread of the disease in our population.

\section{Conclusions and Future Work}
This work has shown that the cumulative death count ($D$) is highly sensitive to the death ratio ($\alpha$) and shows increasing sensitivity to the vaccine rollout rate ($\delta$)in the long run. On the other hand, the cumulative infected count ($W$) is highly sensitive to the reciprocal of the latency period ($\kappa$) at the beginning of the simulation, but this sensitivity weakens over time. In contrast, $W$ is not very sensitive to $\delta$ initially but becomes very sensitive to it in the long run. This supports the importance of the vaccine rollout rate over the vaccine efficacy rate in curbing the spread of the disease in the population. It is also worthwhile to reduce the death ratio by developing a cure for the disease or improving the healthcare system as a whole.

Due to the non-monotonic relationship between $W$ and $\alpha$, this case was excluded from the analysis. Another method of sensitivity analysis (that does not rely on the monotonic relationship between input and output variables) may reveal whether $W$ is sensitive to $\alpha$. It may also reveal other sensitivity relationships that were not very evident when using the combination of LHS and PRCC.

This study presents a general framework for conducting sensitivity analysis on an epidemic model incorporating a mass vaccination program, assuming the distributions of the input variables are as presented in Table \ref{tab:inputparams1}. This framework may be extended to the multiple-city version of model (\ref{eq:one-city}). This will be discussed in a future paper, focusing on the availability of multiple vaccine brands being distributed in the population. One could also look at relaxing the condition $\alpha_v=0$ to see if $D$ will be sensitive to $r$.

A limitation of LHS-PRCC in conducting sensitivity analysis is the monotonicity requirement of the model output with respect to the input variables. This may be influenced by the base values of the model that was used in this study, as shown by preliminary work in this direction but with different distributions having much higher base values for $\beta$ and $\alpha$ \cite{teodoro}. Future studies could be done in exploring these possibilities, simulating the arrival of a pandemic with a much higher casualty rate.

\section*{Acknowledgments}
The author would like to thank Dr. Rachelle R. Sambayan, Dr. Jose Ernie C. Lope, and Ms. Hanna Rhae Lyssa D. Improso for their initial work on the model used in this study, and Mr. John Andrei C. Teodoro for the initial work on the sensitivity analysis of the model.

\section*{Conflicts of Interest}
The authors declare that there are no conflicts of interest.

\bibliographystyle{plain}
\bibliography{references}

\end{document}